\documentclass[preprint]{revtex4}
\usepackage{graphicx,subfigure}
\usepackage{amsfonts}
\usepackage{epsfig}
\usepackage[usenames]{color}

\begin{document}

\title{Viscosity, wave damping and shock wave formation in cold hadronic matter}

\author{D.A. Foga\c{c}a\dag\,  F.S. Navarra\dag\ and L.G. Ferreira Filho\ddag\ \S }
\address{\dag\ Instituto de F\'{\i}sica, Universidade de S\~{a}o Paulo\\
 C.P. 66318, 05315-970 S\~{a}o Paulo, SP, Brazil}
\address{\ddag\ Faculdade de Tecnologia, Universidade do Estado do Rio de Janeiro \\
Via Dutra km 298, 27537-000 Resende, RJ, Brazil}
\address{\S Department of Physics and Astronomy, York University\\ Toronto, Ontario, Canada, M3J 1P3.}

\begin{abstract}
We study linear and nonlinear wave propagation  in a dense and cold hadron gas and also in a
cold quark gluon plasma,  taking viscosity into account and using the Navier-Stokes equation.
The equation of state of the  hadronic phase is derived from the nonlinear Walecka model in the
mean field approximation. The quark gluon plasma phase is described by the MIT equation of state.
We show that in a hadron gas   viscosity strongly damps wave propagation and also hinders
shock wave formation. This marked difference between the two phases may have phenomenological consequences
and lead to new QGP signatures.

\end{abstract}

%\pacs{PACS Numbers : 21.65.Mn, 12.38.Mh, 66.20.Cy, 52.35.Tc}
\maketitle

\vspace{1cm}
\section{Introduction}

Before the appearance of the RHIC data, a tremendous effort had been dedicated to the search of quark gluon plasma (QGP)
signatures and to determine ways of experimentally disentangling the QGP from a hadron gas (HG) \cite{signal}.
Among the most promising signatures we could find $J/\psi$
suppression and enhancement of strangeness production. All the proposed signatures were based on medium properties which were very different in each one of
the two phases. Charmonium suppression, for example, was based on the QGP properties of deconfinement and plasma screening \cite{matsui} .
Strangeness enhancement was based on the chiral symmetry restoration, which is fully realized  in QGP but only partially in a hadron gas \cite{muller}.

In the most recent years the search for QGP signatures had lower  priority, mostly because we believe that QGP has been already produced at RHIC and at
LHC and now the focus should be on the study of its properties  \cite{new-qgp}. We believe that it is still interesting to look for QGP signatures and in this
work we take the first steps to propose a new one, based on  viscosity.

Recently  we have  learned many things about the QGP and one of the most striking
is that it is an almost perfect fluid with very small viscosity  \cite{new-qgp,qgp-visc}. We have also learned, albeit with much larger uncertainties,
that the hadron gas has a very large viscosity \cite{visc,viscc}. According to some estimates, the  viscosity coefficients (both bulk and shear)
in the two phases can differ by two orders of magnitude or more!  This difference is so big that it motivates us to look for observables which are sensitive to
it. So far the existing calculations have only addressed the effects of viscosity in the QGP phase and the shear and bulk viscosity coefficients were varied in
a limited range. These changes produce some visible but not very large effects on global observables such as rapidity and transverse momentum distributions, flow
coefficients and particle correlations \cite{visc-eff}.

We will investigate the  viscosity effects on  wave propagation in a quark gluon plasma and also in a hadron gas. These waves may be caused,
for example, by fluctuations in baryon number or energy density. These fluctuations may be  produced  by inhomogeneous initial conditions, which,
as pointed out in \cite{kapu},  are the result of quantum fluctuations in the densities of the two colliding nuclei and also in the energy
deposition mechanism. These fluctuations and their phenomenological implications have  been studied extensively
\cite{shuryak1,shuryak2,flor,peter,iniflu}  because they may be responsible  for the angular correlations of particle emission observed in heavy-ion
experiments. There are also hydrodynamic fluctuations \cite{kapu}, which are the result of finite particle number effects in a given fluid cell.
This generates local thermal fluctuations of the energy density (and flow velocity) which propagate throughout the fluid. Furthermore, there may be
fluctuations induced by  energetic partons, which have been scattered in the initial collision of the two nuclei and propagate through the medium,
loosing energy and  acting as a source term for the hydrodynamical equations. Finally, there may be also freeze-out fluctuations,
which may be caused by finite particle number effects during and after the freeze-out of the hydrodynamically expanding fluid.

According to our current picture of high energy heavy ion collisions,
a dense and hot hadron gas may exist at later stages of the collisions, when the fluid has
cooled down and the quark-hadron phase transition has taken place.  Actually, as pointed out in
\cite{fodor}, in the finite temperature case there is no transition but rather a cross-over.  Since we consider only a pure QGP or a pure HG and do not
study waves in the presence of phase changes, the real nature of the cold quark-hadron transition (which is not know) is not important in our work.
A dense hadron gas may  also exist and play a more important role in lower energy collisions,
such as those which will be performed at FAIR \cite{fair}. Finally, a dense and cold hadron gas is expected to exist in the core of compact stars.
It is not clear which mechanism might be responsible for the production of waves in the hadronic phase, but it is nevertheless necessary to know how waves
propagate in a HG. In particular, we need to know to what extent waves generated in the QGP survive the passage through a hadron gas.

At intermediate energy proton-nucleus reactions, where we do not expect to form QGP, the impinging proton may be absorbed by the nuclear
medium creating a perturbation in baryon density,
which then propagates through the target. This wave may have a relatively large amplitude and must then be treated beyond the linearization
approximation. This was first done in \cite{raha}, where it was found that, for a  perfect fluid and for a particular equation of state, a
Korteweg - de Vries (KdV) soliton is formed. In \cite{fn06} a similar analysis was carried out with an equation of state derived from the nonlinear
Walecka model with the same conclusion.  The formalism was subsequently extended to relativistic hydrodynamics in \cite{fn07}, to the case of
spherical waves in \cite{fn07a} and finally  to a hadron gas at finite temperature in \cite{ffn09},  where a more detailed numerical solution of
the differential equations was performed and a wider class of relativistic mean field models was investigated.  For some equations of state and/or
some approximation schemes we find KdV solitons and for others we do not find them, but even when there is no soliton we always observe that localized
pulses propagate for long distances (from 10 up to 20 fm !) without loosing strength. In \cite{ffn10} and \cite{ffn11} we  completed the studies of
perfect fluids including the quark gluon plasma at zero and finite temperatures. Even for the QGP, a fluid  with a very different equation of state,
the main features of pulse propagation were still very similar to those found in the previous works.  A really strong difference was observed in
\cite{fgn12}  where viscosity was introduced for the first time in our study of waves. In that work we studied the time evolution of cylindrical ``hot
spots'' (or simply ``tubes'')  in a hot and viscous quark gluon plasma. We could clearly see that a small viscosity  strongly damps and broadens
the tubes during their expansion and they are more easily mixed with the background fluid.

In the present work we perform a systematic comparison between
waves (both linear and nonlinear) in a hadron gas and in a quark gluon plasma, focusing on the effects of viscosity.
For simplicity we shall consider only the case where the temperature
is zero and derive the wave equations for perturbations in a non-relativistic viscous fluid. We will then find the solutions of these equations and see how
they change when we change the viscosity coefficients. Our work has some similarities with Refs. \cite{rand1} and \cite{rand2}, where the effects of
density fluctuations and viscosity on phase separation in heavy ion collisions were investigated. Along this same line, in Refs. \cite{shokov1} and \cite{kapusta}
it was shown that viscosity plays a major role during the nucleation of a hadron bubble in a QGP and also in the nucleation of a QGP bubble in a hadron gas.
The nucleation rate of the forming  phase is directly proportional to the viscosity coefficients of the phase in which the system is. All these works address phase
coexistence whereas we will discuss only pure phases. It is interesting to observe that here and in the mentioned works viscosity plays a  crucial role.

In the framework of non-relativistic fluid dynamics we can derive a theory of dissipative fluids which is remarkably successful: the Navier-Stokes (NS)
theory \cite{roma,wein,land}.  We can use this theory to investigate the evolution of density perturbations  in a hadron gas and in a quark gluon plasma.
Perturbations  are usually studied with the linearization formalism \cite{roma,hidro}, which is the simplest way to study small deviations from equilibrium
and to obtain a wave equations.  The propagation of perturbations through a QGP  has been investigated in several works with the help of  a linearized
version of the hydrodynamics of perfect fluids \cite{shuryak1,shuryak2}  and also of viscous fluids \cite{flor}.
We will later go beyond  linearization and study the effects of viscosity on the propagation of nonlinear waves. To this end we employ the well established
reductive perturbation method (RPM) \cite{davidson,rpm,leblond,loke,nrev,fng}.

This work is organized as  follows. In the next section we present the non-relativistic viscous fluid dynamics. In  section III  we briefly review
the main expressions of the equation of state  of both  the  hadron gas and the quark gluon plasma  at zero temperature.
In  section IV  we combine the hydrodynamical equations with the  equations of state
and derive the  wave equations and present a  numerical study.  The last section is devoted to some concluding remarks.

\section{Non-relativistic Viscous Fluid Dynamics}

In what follows we employ the natural units $c=1$ and $\hbar=1$.  The metric is $g^{\mu\nu}=\textrm{diag}(+,-,-,-)$.
The non-relativistic Navier-Stokes equation is given by
\cite{roma,wein,land}:
\begin{equation}
{\frac{\partial v^{i}}{\partial t}} +v^{k} {\frac{\partial v^{i}}{\partial x^{k}}}=
-{\frac{1}{\rho}}{\frac{\partial p}{\partial x^{i}}}-{\frac{1}{\rho}}{\frac{\partial \Pi^{ki}}{\partial x^{k}}}
\label{ns}
\end{equation}
with the viscous tensor given by:
\begin{equation}
\Pi^{ki}=-\eta\Bigg({\frac{\partial v^{i}}{\partial {x^{k}}}}
+{\frac{\partial v^{k}}{\partial {x^{i}}}}-{\frac{2}{3}}\delta^{ki}
{\frac{\partial v^{l}}{\partial {x^{l}}}}  \Bigg)
-\zeta \delta^{ki}{\frac{\partial v^{l}}{\partial {x^{l}}}}
\label{vt}
\end{equation}
The viscous coefficients are $\eta$ for the shear viscosity and $\zeta$ for the bulk viscosity.
We shall consider $(i=k=l=x)$ for the one dimensional cartesian case, so the Navier-Stokes equation (\ref{ns}) with (\ref{vt}) becomes:
\begin{equation}
{\frac{\partial v_{x}}{\partial t}} +v_{x} {\frac{\partial v_{x}}{\partial x}}=
-{\frac{1}{\rho}}{\frac{\partial p}{\partial x}}+{\frac{1}{\rho}}
\Bigg(\zeta + {\frac{4}{3}}\eta \Bigg){\frac{\partial^{2} v_{x}}{\partial x^{2}}}
\label{nsx}
\end{equation}
where $\rho$ is the mass density. The continuity equation for the mass density $\rho$ is given by:
\begin{equation}
{\frac{\partial \rho}{\partial t}} + {{\nabla}} \cdot (\rho {\vec{v}})=0
\label{cont1}
\end{equation}
In the  nuclear medium, the mass density is
related to the baryon density through $\rho=M\rho_{B}$,
where $M$ is the nucleon mass. The continuity equation for the baryon density
in one cartesian dimension is:
\begin{equation}
{\frac{\partial \rho_{B}}{\partial t}} + v_{x}{\frac{\partial \rho_{B}}{\partial x}}+\rho_{B}{\frac{\partial v_{x}}{\partial x}}=0
\label{contbx}
\end{equation}

In principle the viscosity coefficients, $\eta$ and $\zeta$ may depend on the baryon density.  However, to the best of our knowledge, the dependence of these coefficients (specially in the hadronic phase) with the baryon density is not well known.  In order to compensate for this lack of information
we will consider different values of these coefficients to have an idea of how they affect the wave propagation.

\section{Equation of state}

\subsection{Hadron gas}

The theory of nuclear matter has experienced a continuous progress. Recent developments include the use of effective field theory with chiral power
counting (see for example  \cite{lac}). For our purposes it is enough to work with a theory which reproduces the main features of dense nuclear matter.
In this section we review the derivation of  the equation of state from the nonlinear Walecka  model \cite{furn,serot,fuku} of cold nuclear matter.
The Lagrangian density  is given by:
$$
\mathcal{L}=\bar{\psi}[\gamma_{\mu}(i \partial^{\mu} - g_{V}V^{\mu})-(M-g_{S} \phi)]\psi +
{\frac{1}{2}}\Big(\partial_{\mu} \phi \partial^{\mu} \phi - {m_{S}}^{2} \phi^{2}\Big)
-{\frac{1}{4}}F_{\mu \nu}F^{\mu \nu}+
$$
\begin{equation}
+{\frac{1}{2}}{m_{V}}^{2}V_{\mu}V^{\mu}-{\frac{b}{3}}\phi^{3}-{\frac{c}{4}}\phi^{4}
\label{lagra}
\end{equation}
where $F_{\mu \nu} =  \partial_{\mu} V_{\nu} - \partial_{\nu} V_{\mu}$.
The degrees of freedom are the baryon field $\psi$, the neutral scalar meson field $\phi$ and the neutral vector meson field $V_{\mu}$, with the
respective couplings and masses. The equation of state is obtained through the usual mean field theory (MFT) approximation which consists in considering
the meson fields as classical fields \cite{furn,serot,fuku}: $V_{\mu} \rightarrow <V_{\mu}> \equiv \delta_{\mu 0} \,V_{0}$ and
$\phi \rightarrow <\phi> \equiv \phi_{0}$.  This classical approach is based on the assumption that the baryonic sources are intense,
their coupling to the meson fields is strong and the infinite nuclear matter is static, homogeneous  and isotropic. From the calculations performed in
\cite{furn,serot,fuku} we have the following equations of motion in MFT:
\begin{equation}
{m_{V}}^{2}V_{0}=g_{V}\psi^{\dagger} \psi
\label{veap}
\end{equation}
\begin{equation}
{m_{S}}^{2} \phi_{0}=
g_{S}\bar{\psi}{\psi}-b{\phi_{0}}^{2}-c{\phi_{0}}^{3}
\label{fiap}
\end{equation}
\begin{equation}
\bigg[ i\gamma_{\mu}\partial^{\mu}-g_{V}\gamma_{0}V_{0}
-(M-g_{S} \phi_{0})\bigg]\psi=0
\label{psiap}
\end{equation}
The baryon density, $\rho_{B}$,   is given by:
\begin{equation}
\psi^{\dagger} \psi \equiv \rho_{B}={\frac{\gamma_s}{6 \pi^{2}}}{k_{F}}^{3}
\label{bard}
\end{equation}
where $k_F$ is the Fermi momentum. The equation for the vector meson (\ref{veap}) gives $V_{0}=g_{V}\rho_{B}/m_{V}$.
Equation (\ref{psiap}) is the Dirac equation which couples  the nucleons to the vector mesons. It
gives the fermion contribution (described by the integral term) to the energy density, which is given by:
\begin{equation}
\varepsilon={\frac{{g_{V}}^{2}}{2{m_{V}}^{2}}}{\rho_{B}}^{2}+
{\frac{{m_{S}}^{2}}{2{g_{S}}^{2}}}(M-M^{*})^{2}+
b{\frac{(M-M^{*})^{3}}{3{g_{S}}^{3}}}+
c{\frac{(M-M^{*})^{4}}{4{g_{S}}^{4}}}
+{\frac{\gamma_s}{(2\pi)^{3}}}\int_{0}^{k_{F}} d^3{k}\hspace{0.2cm}\sqrt{{\vec{k}}^{2}+{M^{*}}^{2}}
\label{preeps}
\end{equation}
where $\gamma_s=4$ is the nucleon degeneracy factor and the effective mass of the nucleon is defined as $M^{*}=M-g_{S} \, \phi_{0}$.
The nucleon effective mass is determined by the self-consistency relation obtained from the minimization  of
$\varepsilon(M^{*})$ with respect to $M^{*}$. From (\ref{preeps}) we have:
\begin{equation}
M^{*}=M-{\frac{{g_{S}}^{2}}{{m_{S}}^{2}}}{\frac{\gamma_{s}}{(2\pi)^{3}}}\int^{k_F}_0 d^3{k}{\frac{M^{*}}{\sqrt{{\vec{k}}^{2}+{M^{*}}^{2}}}}
+{\frac{{g_{S}}^{2}}{{m_{S}}^{2}}}\Bigg[{\frac{b}{{g_{S}}^{3}}}(M-M^{*})^{2}+
{\frac{c}{{g_{S}}^{4}}}(M-M^{*})^{3}\Bigg]
\label{efmass}
\end{equation}
In numerical calculations we use from \cite {furn,serot,fuku,dp} the following values for masses and couplings: $M=939 \, MeV$,
$m_{V}=783 \, MeV$, $m_{S}=550 \, MeV$,  $b = 13.47 \, fm^{-1}$, $g_{V}=9.197$, $g_{S}=8.81$ and $c = 43.127$. For the density
\cite {furn,serot,fuku,dp}: $\rho_{0} \leq \rho_{B} \leq 2\rho_{0}$ where $\rho_{0}=0.17 \, fm^{-3}$ is the nuclear baryon density.
From (\ref{efmass}) and (\ref{bard})  we can see that $M^*$ depends on $\rho_B$. Solving (\ref{efmass}) numerically, parametrizing the solution
$M^* = M^* (\rho_B)$  and inserting it into (\ref{preeps}), the energy density can be  rewritten as the following power series in the
baryon density (see the Appendix for details):
$$
\varepsilon=\Big(0.1{\frac{{m_{S}}^{2}}{{g_{S}}^{2}}}+0.04{\frac{b}{{g_{S}}^{3}}}+0.01{\frac{c}{{g_{S}}^{4}}}\Big)
+\Bigg(4+2{\frac{{m_{S}}^{2}}{{g_{S}}^{2}}}+{\frac{b}{{g_{S}}^{3}}}+0.43{\frac{c}{{g_{S}}^{4}}}\Bigg){\rho_{B}}
$$
$$
+\Bigg(-3.75+{\frac{{g_{V}}^{2}}{2{m_{V}}^{2}}}+8{\frac{{m_{S}}^{2}}{{g_{S}}^{2}}}+7.6{\frac{b}{{g_{S}}^{3}}}+5.42{\frac{c}{{g_{S}}^{4}}}\Bigg){\rho_{B}}^{2}
+\Bigg(21.26{\frac{b}{{g_{S}}^{3}}}+30.35{\frac{c}{{g_{S}}^{4}}}\Bigg){\rho_{B}}^{3}
$$
\begin{equation}
+\Bigg(63.73{\frac{c}{{g_{S}}^{4}}}\Bigg){\rho_{B}}^{4}
-1.22\,{\rho_{B}}^{8/3}
+2.61\,{\rho_{B}}^{5/3}
-1.4\,{\rho_{B}}^{2/3}
\label{aepsmefusedag}
\end{equation}

\subsection{QGP}

The equation of state  of the plasma phase is derived from the MIT bag model which describes a gas of noninteracting quarks and gluons and takes into account
nonperturbative  effects through the bag constant $\mathcal{B}$ \cite{nrev}. This constant is interpreted as the energy
needed to create a bubble (or bag) in the QCD physical vacuum. The baryon density is given by:
\begin{equation}
\rho_{B}={\frac{1}{3}}{\frac{\gamma_Q}{(2\pi)^{3}}}\int d^3{k}\hspace{0.2cm}
[n_{\vec{k}}-\bar{n}_{\vec{k}}]
\label{rodensTq}
\end{equation}
with
\begin{equation}
n_{\vec{k}} \equiv n_{\vec{k}}(T)={\frac{1}{1+e^{(k-{\frac{1}{3}}\mu)/ T}}}
\label{qdis}
\hspace{1.6cm} \textrm{and} \hspace{1.6cm}
\bar{n}_{\vec{k}} \equiv \bar{n}_{\vec{k}}(T)={\frac{1}{1+e^{(k+{\frac{1}{3}}\mu)/ T}}}
\label{aqdis}
\end{equation}
where $\mu$ is the baryon chemical potential. At zero temperature the expression for
the baryon density reduces to:
\begin{equation}
{\rho_{B}}={\frac{2}{3\pi^{2}}}{k_{F}}^{3}
\label{barda}
\end{equation}
and  ${k_{F}}$ refers to highest occupied momentum  level.
The energy density and the pressure are given respectively by:
\begin{equation}
\varepsilon=\mathcal{B}+{\frac{\gamma_G}{(2\pi)^{3}}}\int d^3{k}\hspace{0.2cm}k\hspace{0.2cm}(e^{k/T}-1)^{-1}
+{\frac{\gamma_Q}{(2\pi)^{3}}}\int d^3{k}\hspace{0.2cm}k\hspace{0.2cm}[n_{\vec{k}}
+\bar{n}_{\vec{k}}]
\label{esdTqg}
\end{equation}
and
\begin{equation}
p=-\mathcal{B}+{\frac{1}{3}}\Bigg\lbrace {\frac{\gamma_G}{(2\pi)^{3}}}\int d^3{k}\hspace{0.2cm}k\hspace{0.2cm}(e^{k/T}-1)^{-1}+
{\frac{\gamma_Q}{(2\pi)^{3}}}\int d^{3}k\hspace{0.1cm}k\bigg[{n}_{\vec{k}}+\bar{n}_{\vec{k}}\bigg] \Bigg\rbrace
\label{psdTqg}
\end{equation}
The statistical factors  are
$\gamma_G=2\textrm{(polarizations)}\times 8\textrm{(colors)}=16$ for gluons and
$\gamma_Q=2\textrm{(spins)}\times 2\textrm{(flavors)}\times 3
\textrm{(colors)}=12$ for quarks. From the above expressions we
extract the speed of sound $c_{s}$:
\begin{equation}
{c_{s}}^{2}={\frac{\partial p}{\partial \varepsilon}}={\frac{1}{3}}
\label{soundone}
\end{equation}
For the cold plasma at zero temperature and high baryon density the quark distribution functions become  step functions.
Inserting (\ref{barda}) into (\ref{esdTqg}) and into (\ref{psdTqg}) we find
the energy density and pressure as functions of the baryon density, respectively:
\begin{equation}
\varepsilon(\rho_{B})=\bigg({\frac{3}{2}}\bigg)^{7/3}\pi^{2/3}{\rho_{B}}^{4/3}
+\mathcal{B}
\label{epsp}
\end{equation}
and
\begin{equation}
p(\rho_{B})={\frac{1}{3}}\bigg({\frac{3}{2}}\bigg)^{7/3}\pi^{2/3}
{\rho_{B}}^{4/3}-\mathcal{B}
\label{pp}
\end{equation}

\section{Wave  equations}

\subsection{Linear waves}

Before addressing the nonlinear wave equations, we revisit the linear wave equation approach in viscous hydrodynamics.
The non-relativistic Navier-Stokes equation may provide a simple wave equation for
perturbations in the baryon density based on the linearization formalism \cite{roma,hidro}. The use of this formalism can be
justified when small perturbations are considered in a fluid at rest with  background pressure $p_{0}$ and background baryon
density ${\rho_{0}}$.  For simplicity we consider  perturbations which  depend only on the $x$ space coordinate and on time.
We first define the dimensionless variables for the baryon density, for the fluid velocity and for the pressure:
\begin{equation}
\hat\rho(x,t)={\frac{\rho_{B}(x,t)}{\rho_{0}}} \hspace{0.3cm} \textrm{,}   \hspace{1cm}
{\hat v}_{x}(x,t)={\frac{v_{x}(x,t)}{c_{s}}} \hspace{1cm} \textrm{and} \hspace{1cm}  \hat p(x,t)={\frac{p(x,t)}{p_{0}}}
\label{hatvar}
\end{equation}
The expression for the  speed of sound, $c_{s}$, is:
\begin{equation}
{c_{s}}^{2}={\frac{\partial p}{\partial \varepsilon}}
\hspace{0.6cm} \Rightarrow \hspace{0.6cm} p \sim {c_{s}}^{2} \varepsilon
\label{eos}
\end{equation}
In the non-relativistic limit $\varepsilon \cong \rho$,  where $\rho$ is the matter density, which is related to
the baryon density through $\rho=M\rho_{B}$.  We can then write:
\begin{equation}
p = {c_{s}}^{2} M\rho_{B}
\label{eosa}
\end{equation}
Rewriting (\ref{ns}), (\ref{contbx}) and (\ref{eosa}) in one cartesian dimension and using (\ref{hatvar}) we find:
\begin{equation}
M{\rho_{0}}\,{\hat\rho}\Bigg({c_{s}}{\frac{\partial {\hat v}_{x}}{\partial t}}+{c_{s}}^{2}\,
{\hat v}_{x}{\frac{\partial {\hat v}_{x}}{\partial x}}\Bigg)+p_{0}{\frac{\partial {\hat p}}{\partial x}}=
\bigg(\zeta+{\frac{4}{3}}\eta \bigg){c_{s}}
{\frac{\partial^{2} {\hat v}_{x}}{\partial x^{2}}}
\label{nsad}
\end{equation}
\begin{equation}
{\frac{\partial {\hat\rho}}{\partial t}}+{c_{s}}{\hat v}_{x}{\frac{\partial {\hat\rho}}{\partial x}}
+{c_{s}}{\hat\rho}{\frac{\partial {\hat v}_{x}}{\partial x}}=0
\label{contbxad}
\end{equation}
and
\begin{equation}
p_{0}{\hat p}=M{\rho_{0}}\,{c_{s}}^{2} {\hat\rho}
\label{eosaad}
\end{equation}
The perturbations are described by the following expansions of the dimensionless variables:
\begin{equation}
\hat\rho=1+\delta\rho_{B}\hspace{0.3cm} \textrm{,}   \hspace{1cm}
{\hat v}_{x}=\delta v_{x} \hspace{1cm} \textrm{and} \hspace{1cm}  {\hat p}=1+\delta p
\label{perts}
\end{equation}
where  $\delta \rho_B$, $\delta v_{x}$  and $\delta p$,  denote small deviation of the baryon density, velocity and pressure from their equilibrium values respectively.
When the expansions (\ref{perts}) are inserted into the relevant equations in consideration, the ``linearization approximation'' is performed by neglecting the
$\mathcal{O}({\delta^{2}})$ terms \cite{roma,hidro}. Inserting (\ref{perts}) into (\ref{nsad}), (\ref{contbxad}), (\ref{eosaad}) and performing the linearization,
we find respectively:
\begin{equation}
M{\rho_{0}}\,{c_{s}}{\frac{\partial}{\partial t}}\delta v_{x}+
p_{0}{\frac{\partial}{\partial x}}\delta p=\bigg(\zeta+{\frac{4}{3}}\eta \bigg)
{c_{s}}{\frac{\partial^{2}}{\partial x^{2}}}\delta v_{x}
\label{nonrnsagainnfinall}
\end{equation}
\begin{equation}
{\frac{\partial }{\partial t}}\delta\rho_{B}+
{c_{s}}{\frac{\partial}{\partial x}}\delta v_{x}=0
\label{nonrelrhobconsml}
\end{equation}
and
\begin{equation}
p_{0}(1+\delta p)=M{\rho_{0}}\,{c_{s}}^{2}(1+\delta\rho_{B})
\label{eosal}
\end{equation}
Differentiating (\ref{nonrelrhobconsml}) with respect to $x$ and inserting it into
(\ref{nonrnsagainnfinall}) we find:
\begin{equation}
M{\rho_{0}}\,{c_{s}}{\frac{\partial}{\partial t}}\delta v_{x}+
p_{0}{\frac{\partial}{\partial x}}\delta p=-\bigg(\zeta+{\frac{4}{3}}\eta \bigg)
{\frac{\partial}{\partial t}}{\frac{\partial }{\partial x}}\delta\rho_{B}
\label{nssubst}
\end{equation}
Performing the derivative of (\ref{nssubst}) with respect to $x$ and using (\ref{eosal}) , we find:
\begin{equation}
M{\rho_{0}}\,{c_{s}}{\frac{\partial}{\partial t}}{\frac{\partial}{\partial x}}\delta v_{x}+
M{\rho_{0}}\,{c_{s}}^{2}{\frac{\partial^{2}}{\partial x^{2}}}\delta\rho_{B}=-\bigg(\zeta+{\frac{4}{3}}\eta \bigg)
{\frac{\partial}{\partial t}}{\frac{\partial^{2} }{\partial x^{2}}}\delta\rho_{B}
\label{nssubstweos}
\end{equation}
Inserting the time derivative of (\ref{nonrelrhobconsml}) into (\ref{nssubstweos})
we obtain the following viscous wave equation for the baryonic density perturbation
\cite{osalemao}:
\begin{equation}
{\frac{\partial^{2}}{\partial x^{2}}}\delta\rho_{B}
-{\frac{1}{{c_{s}}^{2}}}{\frac{\partial^{2}}{\partial t^{2}}}\delta\rho_{B}=
- {\frac{\nu}{{c_s}^{2}}}   {\frac{\partial}{\partial t}}{\frac{\partial^{2} }{\partial x^{2}}}\delta\rho_{B}
\label{waveeqnr}
\end{equation}
where
\begin{equation}
\nu = {\frac{1}{M{\rho_{0}}}}\bigg(\zeta+{\frac{4}{3}}\eta \bigg)
\label{viscoefdalembert}
\end{equation}
and making use of the Ansatz:
\begin{equation}
\delta \rho_B(x,t)=\mathcal{A}e^{ikx-i\omega t}
\label{plane}
\end{equation}
in (\ref{waveeqnr}) we find the following dispersion relation $\omega(k)$:
\begin{equation}
\omega^{2}={c_{s}}^{2}k^{2}-i\nu \omega k^{2}
\label{dispre}
\end{equation}
where $\nu >0$  and $k \in \mathbb{R}$.  The angular frequency is decomposed in two components \cite{osalemao}:
\begin{equation}
\omega=\omega_{R}+i \omega_{I}
\label{omegaddiv}
\end{equation}
where $\omega_{R} \in \mathbb{R}$ and $\omega_{I} \in \mathbb{R}$.
Inserting (\ref{omegaddiv}) in (\ref{dispre}) we find:
\begin{equation}
{\omega_{R}}^{2}={c_{s}}^{2}k^{2}-{\frac{{\nu}^{2}k^{4}}{4}}
\hspace{2cm} \textrm{and} \hspace{2cm}
{\omega_{I}}=-{\frac{{\nu}k^{2}}{2}} < 0
\label{omegacomps}
\end{equation}
From the above we draw two important conclusions concerning the velocities:

$i)$ The phase velocity $v_{p}$ is given by:
\begin{equation}
{v_{p}}^{2}=\Bigg({\frac{{\omega_{R}}}{k}}\Bigg)^{2} = {c_{s}}^{2}-\Bigg({\frac{{\nu}k}{2}}\Bigg)^{2}
\label{vfase}
\end{equation}
We can observe that one of the effects of viscosity is to reduce $v_p$ .

$ii)$ The group velocity $v_{g}$ is calculated using $|\omega|=c_{s}k$ :
\begin{equation}
{v_{g}}={\frac{d|\omega|}{dk}}=c_{s}
\label{vgroup}
\end{equation}
Inserting (\ref{omegacomps}) into (\ref{omegaddiv}) and the resulting value of $\omega$ into the
Ansatz (\ref{plane}), we find the  solution of (\ref{waveeqnr}):
\begin{equation}
\delta\rho_{B}(x,t)=\mathcal{A}e^{-\nu k^{2}t/2}e^{i\big(kx-t\sqrt{{c_{s}}^{2}k^{2}-{\nu}^{2}k^{4}/4}\big)}
\label{deltapdamped}
\end{equation}
with
\begin{equation}
\Re\Big\{\delta\rho_{B}(x,t)\Big\}=\mathcal{A}e^{-\nu k^{2}t/2}cos\Bigg(kx-\sqrt{{c_{s}}^{2}k^{2}-{\frac{{\nu}^{2}k^{4}}{4}}}\, t\Bigg)
\label{deltapdampedtrig}
\end{equation}
Since in the derivation of the above equation we have only used the general expression  (\ref{eos}), it is valid both for the QGP and HG phases.  The
differences  appear because the constants $\nu$ and $c_s$ are different. The former varies over a much wider range of values whereas the latter is never
very different from $1/3$. For the sake of simplicity we will choose a constant  speed of sound ${c_{s}}^{2}=1/3$ for the two phases and all the differences
will come from $\nu$. From (\ref{viscoefdalembert}) we see that the  changes in  density and viscosity are correlated in such a way that in a hadron gas $\eta$
and $\zeta$ are large and $\rho_0$ is relatively small with $\nu$ being large, whereas in the plasma phase the opposite happens and $\nu$ is small.
As an example we study the  effects of viscosity on the waves propagating in a
cold and dense  system. In this case the perturbation $\delta\rho_{B}$ is a deviation of the baryon density from the background density  ${\rho_{0}}$.

In numerical calculations the viscosity coefficients $\zeta$ and $\eta$ may be taken from \cite{visc,viscc} and used in (\ref{viscoefdalembert}).
Using the exact values of these coefficients is not really crucial for our purposes.
It is enough to know that  $\eta$ is dramatically different in different phases.
The phase velocity must be real and therefore in (\ref{deltapdampedtrig}) we must have:
\begin{equation}
k < \frac{2 c_s}{\nu}
\label{cond-k}
\end{equation}
From this inequality we see that, since $\nu_{QGP} << \nu_{HG}$ the maximum value of $k$ is much smaller in the HG than in the QGP and consequently
in a hadron gas there are much less modes than in a quark gluon plasma.
From (\ref{deltapdampedtrig}) we see that for a hadron gas with high viscosity there is a strong damping of the wave  due to the  exponential factor,
whereas in the plasma the oscillations persist for a very long time.

We close this section emphasizing that waves produced by  small perturbations, those which can be treated with the linearization approximation,
behave quite differently in a hadron gas and in a quark gluon plasma. This pronounced difference comes essentially from the very different
viscosity coefficients in these two phases.
In a hadron gas waves are damped and stalled, whereas in a quark gluon plasma they can propagate nearly undisturbed.

Since there are many possible sources of perturbations and since they do not have
to be necessarily very small, this study has to be extended to stronger perturbations and the resulting nonlinear waves.
This is the subject of the next section.

\subsection{Nonlinear waves in a hadron gas}

We start recalling  the relation between the mass density and baryon density, $\rho=M \rho_{B}$, which used in equation (\ref{nsx}) yields:
\begin{equation}
\rho_{B}\Bigg({\frac{\partial v_{x}}{\partial t}} +v_{x} {\frac{\partial v_{x}}{\partial x}}\Bigg)=
-{\frac{1}{M}}{\frac{\partial p}{\partial x}}+{\frac{1}{M}}
\Bigg(\zeta + {\frac{4}{3}}\eta \Bigg){\frac{\partial^{2} v_{x}}{\partial x^{2}}}
\label{nsxa}
\end{equation}
From  the first law of thermodynamics at zero temperature we have:
\begin{equation}
d\varepsilon=\mu_{B} \, d\rho_{B}
\hspace{1.6cm}
\textrm{where}
\hspace{1.6cm}
\mu_{B}={\frac{\partial \varepsilon}{\partial \rho_{B}}}
\label{deps}
\end{equation}
Inserting (\ref{deps}) into the Gibbs relation at zero temperature:
$$
d\varepsilon + dp=\rho_{B} \,d\mu_{B}+\mu_{B}\, d\rho_{B}
$$
we find
$$
dp=\rho_{B} \, d\mu_{B}
$$
and then finally:
\begin{equation}
dp=\rho_{B} \, d\Bigg({\frac{\partial \varepsilon}{\partial \rho_{B}}} \Bigg)
\hspace{1cm} \textrm{and consequently} \hspace{1cm} {\frac{\partial p}{\partial x}}=
\rho_{B} \, {\frac{\partial }{\partial x}}\Bigg({\frac{\partial \varepsilon}{\partial \rho_{B}}}\Bigg)
\label{dpmub}
\end{equation}
Inserting (\ref{dpmub}) into  (\ref{nsxa}) we find:
\begin{equation}
\rho_{B}\Bigg({\frac{\partial v_{x}}{\partial t}} +v_{x} {\frac{\partial v_{x}}{\partial x}}\Bigg)=
-{\frac{1}{M}}\rho_{B} \, {\frac{\partial }{\partial x}}\Bigg({\frac{\partial \varepsilon}{\partial \rho_{B}}}\Bigg)+{\frac{1}{M}}
\Bigg(\zeta + {\frac{4}{3}}\eta \Bigg){\frac{\partial^{2} v_{x}}{\partial x^{2}}}
\label{nsxaa}
\end{equation}
where the relevant quantity  which carries  information about the equation of state is  $\partial \varepsilon/\partial \rho_{B}$, which can be computed
from (\ref{aepsmefusedag}).
We obtain:
$$
{\frac{\partial \varepsilon}{\partial \rho_{B}}}=
\Bigg(254.92{\frac{c}{{g_{S}}^{4}}}\Bigg){\rho_{B}}^{3}
+\Bigg(63.78{\frac{b}{{g_{S}}^{3}}}+91.05{\frac{c}{{g_{S}}^{4}}}\Bigg){\rho_{B}}^{2}
$$
$$
+\Bigg(-7.5+{\frac{{g_{V}}^{2}}{{m_{V}}^{2}}}+16{\frac{{m_{S}}^{2}}{{g_{S}}^{2}}}+15.2{\frac{b}{{g_{S}}^{3}}}
+10.84{\frac{c}{{g_{S}}^{4}}}\Bigg){\rho_{B}}
-3.25\,{\rho_{B}}^{5/3}+4.35\,{\rho_{B}}^{2/3}-{\rho_{B}}^{-1/3}
$$
\begin{equation}
+\Bigg(4+2{\frac{{m_{S}}^{2}}{{g_{S}}^{2}}}+{\frac{b}{{g_{S}}^{3}}}+0.43{\frac{c}{{g_{S}}^{4}}}\Bigg)
\label{depsrho}
\end{equation}

Inserting (\ref{depsrho}) into (\ref{nsxaa}) we find:
$$
\rho_{B}\Bigg({\frac{\partial v_{x}}{\partial t}} +v_{x} {\frac{\partial v_{x}}{\partial x}}\Bigg)=-{\frac{1}{M}}\Bigg[\Bigg(-7.5+{\frac{{g_{V}}^{2}}{{m_{V}}^{2}}}
+16{\frac{{m_{S}}^{2}}{{g_{S}}^{2}}}+15.2{\frac{b}{{g_{S}}^{3}}}+10.84{\frac{c}{{g_{S}}^{4}}}\Bigg)
{\rho_{B}}{\frac{\partial{\rho_{B}}}{\partial x}}
$$
$$
+\Bigg(127.56{\frac{b}{{g_{S}}^{3}}}+182.1{\frac{c}{{g_{S}}^{4}}}\Bigg)
{\rho_{B}}^{2}\,{\frac{\partial{\rho_{B}}}{\partial x}}
+\Bigg(764.76{\frac{c}{{g_{S}}^{4}}}\Bigg){\rho_{B}}^{3}\,{\frac{\partial{\rho_{B}}}{\partial x}}
$$
\begin{equation}
-5.42\,{\rho_{B}}^{5/3}\,{\frac{\partial{\rho_{B}}}{\partial x}}
+2.9\,{\rho_{B}}^{2/3}\,{\frac{\partial{\rho_{B}}}{\partial x}}
+0.33\,{\rho_{B}}^{-1/3}\,{\frac{\partial{\rho_{B}}}{\partial x}}
\Bigg]+{\frac{1}{M}}\Bigg(\zeta + {\frac{4}{3}}\eta \Bigg){\frac{\partial^{2} v_{x}}{\partial x^{2}}}
\label{nsxaahp}
\end{equation}
which is the Navier-Stokes equation for the hadron phase.

We next combine  (\ref{nsxaahp}) and  (\ref{contbx}) to obtain a nonlinear wave equation.  In dealing with these equations we use the
reductive perturbation method (RPM) \cite{davidson,rpm,leblond,loke,nrev,fng}, which is a technique which  allows us to go beyond linearization and
to preserve nonlinearities, dissipative and dispersive terms in the wave equations. We have already used the RPM in our previous works
to study nonlinear waves in relativistic and non-relativistic ideal hydrodynamics  and also to study the evolution of flux in
relativistic viscous hydrodynamics \cite{fgn12}.  A comprehensive review of the technique and its applications in hydrodynamics of strongly interacting fluids can be found in \cite{nrev}.

The background  baryon density is $\rho_{0}$, upon which the perturbation propagates.
We now rewrite (\ref{contbx}) and (\ref{nsxaahp}) using (\ref{hatvar}) to obtain:
\begin{equation}
{\frac{\partial \hat\rho}{\partial t}} + c_{s}{\hat v}_{x}{\frac{\partial \hat\rho}{\partial x}}+c_{s}\hat\rho{\frac{\partial {\hat v}_{x}}{\partial x}}=0
\label{contbxh}
\end{equation}
and
$$
\hat{\rho}\Bigg({\frac{\partial {\hat v}_{x}}{\partial t}} +c_{s}{\hat v}_{x} {\frac{\partial {\hat v}_{x}}{\partial x}}\Bigg)={\frac{{\rho_{0}}}{Mc_{s}}}
{\Bigg(7.5-{\frac{{g_{V}}^{2}}{{m_{V}}^{2}}}
-16{\frac{{m_{S}}^{2}}{{g_{S}}^{2}}}-15.2{\frac{b}{{g_{S}}^{3}}}-10.84{\frac{c}{{g_{S}}^{4}}}\Bigg)
\hat\rho\,{\frac{\partial{\hat\rho}}{\partial x}}}
$$
$$
-{\frac{{\rho_{0}}^{2}}{Mc_{s}}}\Bigg(127.56{\frac{b}{{g_{S}}^{3}}}+182.1{\frac{c}{{g_{S}}^{4}}}\Bigg)
{\hat\rho}^{2}\,{\frac{\partial{\hat\rho}}{\partial x}}
-{\frac{{\rho_{0}}^{3}}{Mc_{s}}}\Bigg(764.76{\frac{c}{{g_{S}}^{4}}}\Bigg){\hat\rho}^{3}
\,{\frac{\partial{\hat\rho}}{\partial x}}
$$
\begin{equation}
+{\frac{5.42\,{\rho_{0}}^{5/3}}{Mc_{s}}}{\hat\rho}^{5/3}
\,{\frac{\partial{\hat\rho}}{\partial x}}
-{\frac{2.9\,{\rho_{0}}^{2/3}}{Mc_{s}}}{\hat\rho}^{2/3}
\,{\frac{\partial{\hat\rho}}{\partial x}}
-{\frac{0.33\,{\rho_{0}}^{-1/3}}{Mc_{s}}}{\hat\rho}^{-1/3}
\,{\frac{\partial{\hat\rho}}{\partial x}}
+{\frac{1}{M\rho_{0}}}\Bigg(\zeta + {\frac{4}{3}}\eta \Bigg){\frac{\partial^{2} {\hat v}_{x}}{\partial x^{2}}}
\label{nsxlendeosh}
\end{equation}
Following \cite{davidson,rpm,leblond,loke}, we  define the ``stretched coordinates'':
\begin{equation}
\xi=\sigma^{1/2}{\frac{(x-{c_{s}}t)}{L}}
\hspace{2cm} \textrm{and}   \hspace{2cm}
\tau=\sigma^{3/2}{\frac{{c_{s}}t}{L}}
\label{xitau}
\end{equation}
where $L$ is a characteristic length scale of the problem (typically the radius of a heavy ion) and $\sigma$ is a small ($0 < \sigma < 1$) expansion parameter.
We also perform the following transformation of the  viscosity coefficients \cite{stvis1,stvis2}:
\begin{equation}
\zeta=\sigma^{1/2} \, \tilde{\zeta} \hspace{2cm} \textrm{and also}  \hspace{2cm}   \eta=\sigma^{1/2} \, \tilde{\eta}
\label{stv}
\end{equation}
From the stretched coordinates (\ref{xitau}) we have the operators:
\begin{equation}
{\frac{\partial}{\partial x}}={\frac{\sigma^{1/2}}{L}}{\frac{\partial}{\partial \xi}}
\hspace{0.5cm}, \hspace{1.0cm}
{\frac{\partial^{2}}{\partial x^{2}}}={\frac{\sigma}{L^{2}}}{\frac{\partial^{2}}{\partial \xi^{2}}}
\hspace{1.0cm} \textrm{and} \hspace{1.5cm}
{\frac{\partial}{\partial t}}=-c_{s}{\frac{\sigma^{1/2}}{L}}{\frac{\partial}{\partial \xi}}
+c_{s}{\frac{\sigma^{3/2}}{L}}{\frac{\partial}{\partial \tau}}
\label{opers}
\end{equation}
Using (\ref{stv}) and (\ref{opers}) we  transport (\ref{contbxh}) and (\ref{nsxlendeosh}) from the $(x,t)$ space to the  $(\xi,\tau)$ space, obtaining:
\begin{equation}
-{\frac{\partial \hat\rho}{\partial \xi}} + \sigma{\frac{\partial \hat\rho}{\partial \tau}}
+{\hat v}_{x}{\frac{\partial \hat\rho}{\partial \xi}}+\hat\rho{\frac{\partial {\hat v}_{x}}{\partial \xi}}=0
\label{contbxhxitau}
\end{equation}
and
$$
\hat\rho\Bigg(-{\frac{\partial {\hat v}_{x}}{\partial \xi}}
+\sigma{\frac{\partial {\hat v}_{x}}{\partial \tau}} +
{\hat v}_{x} {\frac{\partial {\hat v}_{x}}{\partial \xi}} \Bigg)={\frac{{\rho_{0}}}{M{c_{s}}^{2}}}
\Bigg(7.5-{\frac{{g_{V}}^{2}}{{m_{V}}^{2}}}
-16{\frac{{m_{S}}^{2}}{{g_{S}}^{2}}}-15.2{\frac{b}{{g_{S}}^{3}}}-10.84{\frac{c}{{g_{S}}^{4}}}\Bigg)
\hat\rho\,{\frac{\partial{\hat\rho}}{\partial \xi}}
$$
$$
-{\frac{{\rho_{0}}^{2}}{M{c_{s}}^{2}}}\Bigg(127.56{\frac{b}{{g_{S}}^{3}}}+182.1{\frac{c}{{g_{S}}^{4}}}\Bigg){\hat\rho}^{2}
\,{\frac{\partial{\hat\rho}}{\partial \xi}}
-{\frac{{\rho_{0}}^{3}}{M{c_{s}}^{2}}}\Bigg(764.76{\frac{c}{{g_{S}}^{4}}}\Bigg)
{\hat\rho}^{3}\,{\frac{\partial{\hat\rho}}{\partial \xi}}
$$
\begin{equation}
+{\frac{5.42\,{\rho_{0}}^{5/3}}{M{c_{s}}^{2}}}{\hat\rho}^{5/3}
\,{\frac{\partial{\hat\rho}}{\partial \xi}}
-{\frac{2.9\,{\rho_{0}}^{2/3}}{M{c_{s}}^{2}}}{\hat\rho}^{2/3}
\,{\frac{\partial{\hat\rho}}{\partial \xi}}
-{\frac{0.33\,{\rho_{0}}^{-1/3}}{M{c_{s}}^{2}}}{\hat\rho}^{-1/3}
\,{\frac{\partial{\hat\rho}}{\partial \xi}}
+{\frac{\sigma}{M\rho_{0}c_{s}L}}
\Bigg(\tilde{\zeta} + {\frac{4}{3}}\tilde{\eta} \Bigg)
{\frac{\partial^{2} {\hat v}_{x}}{\partial \xi^{2}}}
\label{nsxlendeoshxitau}
\end{equation}
We now  perform the expansion of the dimensionless baryon density and the dimensionless fluid velocity given by (\ref{hatvar})
around their equilibrium values:
\begin{equation}
\hat\rho={\frac{\rho_{B}}{\rho_{0}}}=1+\sigma \rho_{1}+ \sigma^{2} \rho_{2}+
\sigma^{3} \rho_{3}+\dots
\label{barexp}
\end{equation}
\begin{equation}
{\hat v}_{x}={\frac{v_{x}}{c_{s}}}=\sigma {v_{x}}_{1}+ \sigma^{2} {v_{x}}_{2}+ \sigma^{3} {v_{x}}_{3}+\dots
\label{vexp}
\end{equation}
We now insert the expansions (\ref{barexp}) and (\ref{vexp}) into (\ref{contbxhxitau}) and into (\ref{nsxlendeoshxitau}),
neglect the terms proportional to $\sigma^{n}$ for $n > 2$ and organize the equations as series in powers of $\sigma$ and $\sigma^{2}$,
finding the following equations:
\begin{equation}
\sigma\Bigg\{-{\frac{\partial \rho_{1}}{\partial \xi}}+
{\frac{\partial {v_{x}}_{1}}{\partial \xi}}\Bigg\}
+ \sigma^{2}\Bigg\{ -{\frac{\partial \rho_{2}}{\partial \xi}}
+{\frac{\partial \rho_{1}}{\partial \tau}}+ {v_{x}}_{1}{\frac{\partial \rho_{1}}{\partial \xi}}+{\frac{\partial {v_{x}}_{2}}{\partial \xi}}
+\rho_{1}{\frac{\partial {v_{x}}_{1}}{\partial \xi}}\Bigg\}=0
\label{contbxhxitauexp}
\end{equation}
and
$$
\sigma\Bigg\{-{\frac{\partial {v_{x}}_{1}}{\partial \xi}}+
\Bigg[-{\frac{{\rho_{0}}}{M{c_{s}}^{2}}}
\Bigg(7.5-{\frac{{g_{V}}^{2}}{{m_{V}}^{2}}}
-16{\frac{{m_{S}}^{2}}{{g_{S}}^{2}}}-15.2{\frac{b}{{g_{S}}^{3}}}-10.84{\frac{c}{{g_{S}}^{4}}}\Bigg)
$$
$$
+{\frac{{\rho_{0}}^{2}}{M{c_{s}}^{2}}}\Bigg(127.56{\frac{b}{{g_{S}}^{3}}}+182.1{\frac{c}{{g_{S}}^{4}}}\Bigg)
+{\frac{{\rho_{0}}^{3}}{M{c_{s}}^{2}}}\Bigg(764.76{\frac{c}{{g_{S}}^{4}}}\Bigg)
$$
$$
-{\frac{5.42\,{\rho_{0}}^{5/3}}{M{c_{s}}^{2}}}
+{\frac{2.9\,{\rho_{0}}^{2/3}}{M{c_{s}}^{2}}}
+{\frac{0.33\,{\rho_{0}}^{-1/3}}{M{c_{s}}^{2}}}
\Bigg]
{\frac{\partial \rho_{1}}{\partial \xi}}\Bigg\}
$$
\,\,\,
$$
+\sigma^{2}\Bigg\{
-{\frac{\partial {v_{x}}_{2}}{\partial \xi}}
-\rho_{1}{\frac{\partial {v_{x}}_{1}}{\partial \xi}}
+{\frac{\partial {v_{x}}_{1}}{\partial \tau}}+{v_{x}}_{1}{\frac{\partial {v_{x}}_{1}}{\partial \xi}}
+\Bigg[-{\frac{{\rho_{0}}}{M{c_{s}}^{2}}}
\Bigg(7.5-{\frac{{g_{V}}^{2}}{{m_{V}}^{2}}}
-16{\frac{{m_{S}}^{2}}{{g_{S}}^{2}}}-15.2{\frac{b}{{g_{S}}^{3}}}-10.84{\frac{c}{{g_{S}}^{4}}}\Bigg)
$$
$$
+{\frac{{\rho_{0}}^{2}}{M{c_{s}}^{2}}}\Bigg(127.56{\frac{b}{{g_{S}}^{3}}}
+182.1{\frac{c}{{g_{S}}^{4}}}\Bigg)
+{\frac{{\rho_{0}}^{3}}{M{c_{s}}^{2}}}\Bigg(764.76{\frac{c}{{g_{S}}^{4}}}\Bigg)
-{\frac{5.42\,{\rho_{0}}^{5/3}}{M{c_{s}}^{2}}}
+{\frac{2.9\,{\rho_{0}}^{2/3}}{M{c_{s}}^{2}}}
$$
$$
+{\frac{0.33\,{\rho_{0}}^{-1/3}}{M{c_{s}}^{2}}}
\Bigg]\rho_{1}{\frac{\partial {\rho_{1}}}{\partial \xi}}
+\Bigg[-{\frac{{\rho_{0}}}{M{c_{s}}^{2}}}
\Bigg(7.5-{\frac{{g_{V}}^{2}}{{m_{V}}^{2}}}
-16{\frac{{m_{S}}^{2}}{{g_{S}}^{2}}}-15.2{\frac{b}{{g_{S}}^{3}}}-10.84{\frac{c}{{g_{S}}^{4}}}\Bigg)
$$
$$
+{\frac{{\rho_{0}}^{2}}{M{c_{s}}^{2}}}\Bigg(127.56{\frac{b}{{g_{S}}^{3}}}+182.1{\frac{c}{{g_{S}}^{4}}}\Bigg)
+{\frac{{\rho_{0}}^{3}}{M{c_{s}}^{2}}}\Bigg(764.76{\frac{c}{{g_{S}}^{4}}}\Bigg)
-{\frac{5.42\,{\rho_{0}}^{5/3}}{M{c_{s}}^{2}}}
+{\frac{2.9\,{\rho_{0}}^{2/3}}{M{c_{s}}^{2}}}
$$
$$
+{\frac{0.33\,{\rho_{0}}^{-1/3}}{M{c_{s}}^{2}}}
\Bigg]
{\frac{\partial {\rho_{2}}}{\partial \xi}}
+\Bigg[{\frac{{\rho_{0}}^{2}}{M{c_{s}}^{2}}}\Bigg(127.56{\frac{b}{{g_{S}}^{3}}}
+182.1{\frac{c}{{g_{S}}^{4}}}\Bigg)
+{\frac{{\rho_{0}}^{3}}{M{c_{s}}^{2}}}\Bigg(1529.52{\frac{c}{{g_{S}}^{4}}}\Bigg)
$$
\begin{equation}
-{\frac{3.61\,{\rho_{0}}^{5/3}}{M{c_{s}}^{2}}}
-{\frac{0.97\,{\rho_{0}}^{2/3}}{M{c_{s}}^{2}}}
-{\frac{0.44\,{\rho_{0}}^{-1/3}}{M{c_{s}}^{2}}}\Bigg]
\rho_{1}{\frac{\partial {\rho_{1}}}{\partial \xi}}
-{\frac{1}{M\rho_{0}c_{s}L}}\Bigg(\tilde{\zeta} + {\frac{4}{3}}\tilde{\eta} \Bigg)
{\frac{\partial^{2} {{v}_{x}}_{1}}{\partial \xi^{2}}}\Bigg\}=0
\label{nsxlendeoshxitauexp}
\end{equation}
Each bracket in the last two equations must vanish:  $\sigma \{ \dots \}=0$ \,
and \, $\sigma^{2} \{ \dots \}=0$ .  The final step of RPM is collect the equations
from each $\sigma$ order to obtain the nonlinear wave equation.
From the terms proportional to $\sigma$ we have,
in a simple way, after the integration with respect to $\xi$ and setting the integration
constant equals to zero, the following results:
\begin{equation}
{v_{x}}_{1}= \rho_{1}
\label{sigma}
\end{equation}
and the algebraic relation
$$
-{\frac{{\rho_{0}}}{M{c_{s}}^{2}}}
\Bigg(7.5-{\frac{{g_{V}}^{2}}{{m_{V}}^{2}}}
-16{\frac{{m_{S}}^{2}}{{g_{S}}^{2}}}-15.2{\frac{b}{{g_{S}}^{3}}}-10.84{\frac{c}{{g_{S}}^{4}}}\Bigg)
+{\frac{{\rho_{0}}^{2}}{M{c_{s}}^{2}}}\Bigg(127.56{\frac{b}{{g_{S}}^{3}}}
+182.1{\frac{c}{{g_{S}}^{4}}}\Bigg)
$$
\begin{equation}
+{\frac{{\rho_{0}}^{3}}{M{c_{s}}^{2}}}\Bigg(764.76{\frac{c}{{g_{S}}^{4}}}\Bigg)
-{\frac{5.42\,{\rho_{0}}^{5/3}}{M{c_{s}}^{2}}}
+{\frac{2.9\,{\rho_{0}}^{2/3}}{M{c_{s}}^{2}}}
+{\frac{0.33\,{\rho_{0}}^{-1/3}}{M{c_{s}}^{2}}}=1
\label{rela}
\end{equation}
which gives the sound speed $(c_{s})$:
$$
{c_{s}}^{2}=-{\frac{{\rho_{0}}}{M}}
\Bigg(7.5-{\frac{{g_{V}}^{2}}{{m_{V}}^{2}}}
-16{\frac{{m_{S}}^{2}}{{g_{S}}^{2}}}-15.2{\frac{b}{{g_{S}}^{3}}}-10.84{\frac{c}{{g_{S}}^{4}}}\Bigg)
+{\frac{{\rho_{0}}^{2}}{M}}\Bigg(127.56{\frac{b}{{g_{S}}^{3}}}+182.1{\frac{c}{{g_{S}}^{4}}}\Bigg)
$$
\begin{equation}
+{\frac{{\rho_{0}}^{3}}{M}}\Bigg(764.76{\frac{c}{{g_{S}}^{4}}}\Bigg)
-{\frac{5.42\,{\rho_{0}}^{5/3}}{M}}
+{\frac{2.9\,{\rho_{0}}^{2/3}}{M}}
+{\frac{0.33\,{\rho_{0}}^{-1/3}}{M}}
\label{relacs}
\end{equation}
From the terms proportional to $\sigma^{2}$ we find:
\begin{equation}
{\frac{\partial \rho_{2}}{\partial \xi}}-{\frac{\partial {v_{x}}_{2}}{\partial \xi}}=
{\frac{\partial \rho_{1}}{\partial \tau}}+ {v_{x}}_{1}{\frac{\partial \rho_{1}}{\partial \xi}}+ \rho_{1}{\frac{\partial {v_{x}}_{1}}{\partial \xi}}
\label{sigma2a}
\end{equation}
and
$$
{\frac{\partial {v_{x}}_{2}}{\partial \xi}}
-\Bigg[-{\frac{{\rho_{0}}}{M{c_{s}}^{2}}}
\Bigg(7.5-{\frac{{g_{V}}^{2}}{{m_{V}}^{2}}}
-16{\frac{{m_{S}}^{2}}{{g_{S}}^{2}}}-15.2{\frac{b}{{g_{S}}^{3}}}-10.84{\frac{c}{{g_{S}}^{4}}}\Bigg)
+{\frac{{\rho_{0}}^{2}}{M{c_{s}}^{2}}}\Bigg(127.56{\frac{b}{{g_{S}}^{3}}}
+182.1{\frac{c}{{g_{S}}^{4}}}\Bigg)
$$
$$
+{\frac{{\rho_{0}}^{3}}{M{c_{s}}^{2}}}\Bigg(764.76{\frac{c}{{g_{S}}^{4}}}\Bigg)
-{\frac{5.42\,{\rho_{0}}^{5/3}}{M{c_{s}}^{2}}}
+{\frac{2.9\,{\rho_{0}}^{2/3}}{M{c_{s}}^{2}}}
+{\frac{0.33\,{\rho_{0}}^{-1/3}}{M{c_{s}}^{2}}}
\Bigg]{\frac{\partial \rho_{2}}{\partial \xi}}=
{\frac{\partial {v_{x}}_{1}}{\partial \tau}}+{v_{x}}_{1}{\frac{\partial {v_{x}}_{1}}{\partial \xi}}-\rho_{1}{\frac{\partial {v_{x}}_{1}}{\partial \xi}}
$$
$$
+\Bigg\{\,\Bigg[-{\frac{{\rho_{0}}}{M{c_{s}}^{2}}}
\Bigg(7.5-{\frac{{g_{V}}^{2}}{{m_{V}}^{2}}}
-16{\frac{{m_{S}}^{2}}{{g_{S}}^{2}}}-15.2{\frac{b}{{g_{S}}^{3}}}-10.84{\frac{c}{{g_{S}}^{4}}}\Bigg)
+{\frac{{\rho_{0}}^{2}}{M{c_{s}}^{2}}}
\Bigg(127.56{\frac{b}{{g_{S}}^{3}}}
+182.1{\frac{c}{{g_{S}}^{4}}}\Bigg)
$$
$$
+{\frac{{\rho_{0}}^{3}}{M}}\Bigg(764.76{\frac{c}{{g_{S}}^{4}}}\Bigg)
-{\frac{5.42\,{\rho_{0}}^{5/3}}{M}}
+{\frac{2.9\,{\rho_{0}}^{2/3}}{M}}
+{\frac{0.33\,{\rho_{0}}^{-1/3}}{M}}\Bigg]
$$
$$
+\Bigg[{\frac{{\rho_{0}}^{2}}{M{c_{s}}^{2}}}\Bigg(127.56{\frac{b}{{g_{S}}^{3}}}
+182.1{\frac{c}{{g_{S}}^{4}}}\Bigg)
+{\frac{{\rho_{0}}^{3}}{M{c_{s}}^{2}}}\Bigg(1529.52{\frac{c}{{g_{S}}^{4}}}\Bigg)
-{\frac{3.61\,{\rho_{0}}^{5/3}}{M{c_{s}}^{2}}}
-{\frac{0.97\,{\rho_{0}}^{2/3}}{M{c_{s}}^{2}}}
$$
\begin{equation}
-{\frac{0.44\,{\rho_{0}}^{-1/3}}{M{c_{s}}^{2}}}\Bigg]\,\Bigg\}
\rho_{1}{\frac{\partial {\rho_{1}}}{\partial \xi}}
-{\frac{1}{M{\rho_{0}}c_{s}L}}\Bigg(\tilde{\zeta} + {\frac{4}{3}}\tilde{\eta} \Bigg)
{\frac{\partial^{2} {{v}_{x}}_{1}}{\partial \xi^{2}}}
\label{sigma2b}
\end{equation}

Using (\ref{sigma}) in (\ref{sigma2a}) we find:
\begin{equation}
{\frac{\partial \rho_{2}}{\partial \xi}}-{\frac{\partial {v_{x}}_{2}}{\partial \xi}}=
{\frac{\partial \rho_{1}}{\partial \tau}}+ 2\rho_{1}{\frac{\partial \rho_{1}}{\partial \xi}}
\label{sigma2af}
\end{equation}
and using (\ref{sigma}) and (\ref{rela}) in (\ref{sigma2b}) we find:
$$
{\frac{\partial \rho_{2}}{\partial \xi}}-{\frac{\partial {v_{x}}_{2}}{\partial \xi}}=
-{\frac{\partial \rho_{1}}{\partial \tau}}-\rho_{1}{\frac{\partial \rho_{1}}{\partial \xi}}
-\Bigg[{\frac{{\rho_{0}}^{2}}{M{c_{s}}^{2}}}\Bigg(127.56{\frac{b}{{g_{S}}^{3}}}
+182.1{\frac{c}{{g_{S}}^{4}}}\Bigg)
+{\frac{{\rho_{0}}^{3}}{M{c_{s}}^{2}}}\Bigg(1529.52{\frac{c}{{g_{S}}^{4}}}\Bigg)
$$
\begin{equation}
-{\frac{3.61\,{\rho_{0}}^{5/3}}{M{c_{s}}^{2}}}
-{\frac{0.97\,{\rho_{0}}^{2/3}}{M{c_{s}}^{2}}}
-{\frac{0.44\,{\rho_{0}}^{-1/3}}{M{c_{s}}^{2}}}\Bigg]\rho_{1}{\frac{\partial \rho_{1}}{\partial \xi}}
+{\frac{1}{M\rho_{0}c_{s}L}}\Bigg(\tilde{\zeta} + {\frac{4}{3}}\tilde{\eta} \Bigg){\frac{\partial^{2} \rho_{1}}{\partial \xi^{2}}}
\label{sigma2bf}
\end{equation}
Combining (\ref{sigma2af}) and (\ref{sigma2bf}) we find the Burgers equation in the $(\xi,\tau)$ space:
$$
{\frac{\partial \rho_{1}}{\partial \tau}}+
\Bigg\{{\frac{3}{2}}+\Bigg[{\frac{{\rho_{0}}^{2}}{M{c_{s}}^{2}}}\Bigg(127.56{\frac{b}{{g_{S}}^{3}}}
+182.1{\frac{c}{{g_{S}}^{4}}}\Bigg)
+{\frac{{\rho_{0}}^{3}}{M{c_{s}}^{2}}}\Bigg(1529.52{\frac{c}{{g_{S}}^{4}}}\Bigg)
-{\frac{3.61\,{\rho_{0}}^{5/3}}{M{c_{s}}^{2}}}
$$
\begin{equation}
-{\frac{0.97\,{\rho_{0}}^{2/3}}{M{c_{s}}^{2}}}
-{\frac{0.44\,{\rho_{0}}^{-1/3}}{M{c_{s}}^{2}}}\Bigg]{\frac{1}{2}}\Bigg\}
\rho_{1}{\frac{\partial \rho_{1}}{\partial \xi}}
={\frac{1}{2M\rho_{0}c_{s}L}}\Bigg(\tilde{\zeta} + {\frac{4}{3}}\tilde{\eta} \Bigg){\frac{\partial^{2} \rho_{1}}{\partial \xi^{2}}}
\label{kdvBxitau}
\end{equation}
Returning (\ref{kdvBxitau}) to the $(x,t)$ space  (with the help of  (\ref{stv}) and (\ref{opers}) ) we obtain the following Burgers equation:
$$
{\frac{\partial }{\partial t}}\delta\rho_{B}+
{c_{s}}{\frac{\partial }{\partial x}}\delta\rho_{B}
+\Bigg\{{\frac{3}{2}}{c_{s}}+\Bigg[{\frac{{\rho_{0}}^{2}}{M{c_{s}}^{2}}}\Bigg(127.56{\frac{b}{{g_{S}}^{3}}}
+182.1{\frac{c}{{g_{S}}^{4}}}\Bigg)
+{\frac{{\rho_{0}}^{3}}{M{c_{s}}^{2}}}\Bigg(1529.52{\frac{c}{{g_{S}}^{4}}}\Bigg)
$$
\begin{equation}
-{\frac{3.61\,{\rho_{0}}^{5/3}}{M{c_{s}}^{2}}}
-{\frac{0.97\,{\rho_{0}}^{2/3}}{M{c_{s}}^{2}}}
-{\frac{0.44\,{\rho_{0}}^{-1/3}}{M{c_{s}}^{2}}}\Bigg]{\frac{{c_{s}}}{2}}\Bigg\}
\delta\rho_{B} \,
{\frac{\partial }{\partial x}}\delta\rho_{B}=
{\frac{1}{2M\rho_{0}}}\Bigg(\zeta + {\frac{4}{3}}\eta \Bigg){\frac{\partial^{2}}{\partial x^{2}}}\delta\rho_{B}
\label{kdvBxt}
\end{equation}
where  (\ref{barexp}): $\delta\rho_{B}\equiv \sigma \rho_{1}$  is the first deviation from the background  baryon density $\rho_{0}$.

\subsection{Nonlinear waves in a  QGP}

From (\ref{pp}) we have:
\begin{equation}
{\frac{\partial p}{\partial x}}
={\frac{4}{9}}\bigg({\frac{3}{2}}\bigg)^{7/3}\pi^{2/3}{\rho_{B}}^{1/3}
{\frac{\partial \rho_{B}}{\partial x}}
\label{depx}
\end{equation}
In the non-relativistic limit $\varepsilon+p \cong \rho$ and from (\ref{epsp}) and (\ref{pp}) we find:
\begin{equation}
\rho={\frac{4}{3}}\bigg({\frac{3}{2}}\bigg)^{7/3}\pi^{2/3}{\rho_{B}}^{4/3}
\label{epsppre}
\end{equation}
Inserting (\ref{depx}) and (\ref{epsppre}) into (\ref{nsx}) we obtain the following
Navier-Stokes equation for the plasma phase:
\begin{equation}
{\frac{4}{3}}\bigg({\frac{3}{2}}\bigg)^{7/3}\pi^{2/3}{\rho_{B}}^{4/3}\Bigg(
{\frac{\partial v_{x}}{\partial t}} +v_{x} {\frac{\partial v_{x}}{\partial x}}\Bigg)=
-{\frac{4}{9}}\bigg({\frac{3}{2}}\bigg)^{7/3}\pi^{2/3}{\rho_{B}}^{1/3}
{\frac{\partial {\rho_{B}}}{\partial x}}
+\Bigg(\zeta + {\frac{4}{3}}\eta \Bigg){\frac{\partial^{2} v_{x}}{\partial x^{2}}}
\label{nsxppha}
\end{equation}
Repeating all the steps shown in  the last section, we obtain from the terms of order $\sigma$:
\begin{equation}
{v_{x}}_{1}= \rho_{1}  \hspace{1.6cm} \textrm{and}  \hspace{1.6cm}  {c_{s}}^{2}=1/3
\label{sigmag}
\end{equation}
and from terms of  order $\sigma^{2}$:
$$
{\frac{4}{9{c_{s}}^{2}}}\bigg({\frac{3}{2}}\bigg)^{7/3}\pi^{2/3}{\rho_{0}}^{4/3}
{\frac{\partial \rho_{2}}{\partial \xi}}-
{\frac{4}{3}}\bigg({\frac{3}{2}}\bigg)^{7/3}\pi^{2/3}{\rho_{0}}^{4/3}
{\frac{\partial {v_{x}}_{2}}{\partial \xi}}=
{\frac{16}{9}}\bigg({\frac{3}{2}}\bigg)^{7/3}\pi^{2/3}{\rho_{0}}^{4/3}\rho_{1}
{\frac{\partial {v_{x}}_{1}}{\partial \xi}}
$$
$$
-{\frac{4}{27{c_{s}}^{2}}}\bigg({\frac{3}{2}}\bigg)^{7/3}\pi^{2/3}{\rho_{0}}^{4/3}\rho_{1}
{\frac{\partial \rho_{1}}{\partial \xi}}
+{\frac{4}{3}}\bigg({\frac{3}{2}}\bigg)^{7/3}\pi^{2/3}{\rho_{0}}^{4/3}
{\frac{\partial {v_{x}}_{1}}{\partial \tau}}+
{\frac{4}{3}}\bigg({\frac{3}{2}}\bigg)^{7/3}\pi^{2/3}{\rho_{0}}^{4/3}{v_{x}}_{1}
{\frac{\partial {v_{x}}_{1}}{\partial \xi}}
$$
\begin{equation}
+{\frac{1}{L{c_{s}}}}\Bigg(\tilde{\zeta} + {\frac{4}{3}}\tilde{\eta} \Bigg){\frac{\partial^{2} {v_{x}}_{1}}{\partial \xi^{2}}}
\label{sigmaord2}
\end{equation}
Inserting (\ref{sigmag}) into (\ref{sigmaord2}) and combining this result with
(\ref{sigma2af}) we find the following Burgers equation in the $(x,t)$ space:
\begin{equation}
{\frac{\partial }{\partial t}}\delta\rho_{B}+
{c_{s}}{\frac{\partial }{\partial x}}\delta\rho_{B}
+{c_{s}}\, \delta\rho_{B} \,{\frac{\partial }{\partial x}}\delta\rho_{B}=
\Bigg[{\frac{8}{3}}\bigg({\frac{3}{2}}\bigg)^{7/3}\pi^{2/3}{\rho_{0}}^{4/3}\Bigg]^{-1}
\times \Bigg(\zeta + {\frac{4}{3}}\eta \Bigg){\frac{\partial^{2} }{\partial x^{2}}}\delta\rho_{B}
\label{kdvBxtpphase}
\end{equation}
where,  again,  $\delta\rho_{B}\equiv \sigma \rho_{1}$.

Comparing the above expression with (\ref{kdvBxt}) we see that, apart from small numerical diffe-rences in the coefficients, nonlinear waves
obey the same Burgers equation  in both QGP and HG phases. We can thus anticipate that, as it was the case of
linear waves, the most important differences of nonlinear  wave propagation will come from the very different numerical values of the viscosity
coefficients $\eta$ and $\zeta$.

Another equation of state, which is often applied to neutron star physics,  is  given by:
\begin{equation}
\varepsilon(\rho_{B})=\vartheta \, \rho_{B} \hspace{1.2cm} \textrm{and} \hspace{1.2cm}  p(\rho_{B})=\chi \,\rho_{B}
\label{epstif}
\end{equation}
where  $\vartheta$ and $\chi$ are dimension full constants. In the non-relativistic regime, we can derive
from (\ref{epstif})  the following relation:
\begin{equation}
\varepsilon+p\cong \rho =(\vartheta+\chi)\rho_{B}
\label{epstifsum}
\end{equation}
Inserting (\ref{epstif}) and (\ref{epstifsum}) into  the Navier-Stokes equation (\ref{nsx})
and repeating the steps of the RPM formalism  we find the following Burgers equation for $\delta\rho_{B}\equiv \sigma \rho_{1}$:
\begin{equation}
{\frac{\partial }{\partial t}}\delta\rho_{B}+
{c_{s}}{\frac{\partial }{\partial x}}\delta\rho_{B}
+{c_{s}}\, \delta\rho_{B} \,{\frac{\partial }{\partial x}}\delta\rho_{B}=
{\frac{1}{2(\vartheta+\chi)\rho_{0}}}\Bigg(\zeta + {\frac{4}{3}}\eta \Bigg){\frac{\partial^{2} }{\partial x^{2}}}\delta\rho_{B}
\label{kdvBxtpphasestar}
\end{equation}
with
\begin{equation}
{c_{s}}^{2}={\frac{\chi}{(\vartheta+\chi)}}
\label{soundstar}
\end{equation}
Comparing (\ref{kdvBxtpphasestar}) with (\ref{kdvBxtpphase}) we observe that the dissipative term has changed but the equation is the same.
We also have (\ref{soundstar}) instead  of (\ref{sigmag}).

\subsection{Shock waves}

The Burgers equations (\ref{kdvBxt}) and (\ref{kdvBxtpphase}) may be written in the general form:
\begin{equation}
{\frac{\partial }{\partial t}}\delta\rho_{B}+
{c_{s}}{\frac{\partial }{\partial x}}\delta\rho_{B}
+\alpha \, \delta\rho_{B} \,{\frac{\partial }{\partial x}}\delta\rho_{B}=\mu \, {\frac{\partial^{2} }{\partial x^{2}}}\delta\rho_{B}
\label{burgeral}
\end{equation}
where $\alpha$ and $\mu$ are the respective nonlinear and dissipative coefficients for the the  plasma and hadron phase given respectively by:
\begin{equation}
\alpha=
\left\{ \begin{array}{ll}
{\alpha}_{QGP} = c_{s}  \\
{\alpha}_{HG}=
{\frac{3}{2}}{c_{s}}+\Bigg[{\frac{{\rho_{0}}^{2}}{M{c_{s}}^{2}}}\Bigg(127.56{\frac{b}{{g_{S}}^{3}}}
+182.1{\frac{c}{{g_{S}}^{4}}}\Bigg)
+{\frac{{\rho_{0}}^{3}}{M{c_{s}}^{2}}}\Bigg(1529.52{\frac{c}{{g_{S}}^{4}}}\Bigg) \\
\hspace{3cm} -{\frac{3.61\,{\rho_{0}}^{5/3}}{M{c_{s}}^{2}}}
-{\frac{0.97\,{\rho_{0}}^{2/3}}{M{c_{s}}^{2}}}
-{\frac{0.44\,{\rho_{0}}^{-1/3}}{M{c_{s}}^{2}}}\Bigg]{\frac{{c_{s}}}{2}}
\end{array} \right.
\label{nledistermqgphg}
\end{equation}
\begin{equation}
\mu=
\left\{ \begin{array}{ll}
{\mu}_{HG}= {\frac{1}{2M\rho_{0}}} \, \Delta_{HG} \\
{\mu}_{QGP} = \Bigg[{\frac{8}{3}}\bigg({\frac{3}{2}}\bigg)^{7/3}\pi^{2/3}{\rho_{0}}^{4/3}\Bigg]^{-1}
\times \Delta_{QGP}
\end{array} \right.
\label{distermsqgphg}
\end{equation}
where for simplicity we have defined the ``effective viscosity coefficient'' :
\begin{equation}
\Delta =
\left\{ \begin{array}{ll}
{\Delta}_{HG} \equiv \Bigg(\zeta_{HG} + {\frac{4}{3}}\eta_{HG} \Bigg) \\
{\Delta}_{QGP} \equiv \Bigg(\zeta_{QGP} + {\frac{4}{3}}\eta_{QGP} \Bigg)
\end{array} \right.
\label{efectvis}
\end{equation}
which assumes distinct values for QGP and HG phases.

Applying the hyperbolic tangent
function method  as described in \cite{pra1,egypt,pra2} we obtain the exact traveling wave solution of (\ref{burgeral}) with the free parameter $\lambda$:
\begin{equation}
\delta\rho_{B}(x,t;\mu)=-{\frac{2\mu\lambda}{\alpha}}-{\frac{2\mu\lambda}{\alpha}}
tanh\Big\{\lambda \Big[x+(2\mu\lambda-c_{s})t\Big]\Big\}
%\Bigg\{\lambda
%\Bigg[x-\Bigg(c_{s}+{\frac{\alpha}{3}}\Bigg)t\Bigg]\Bigg\}
\label{preburgeralsol}
\end{equation}
For the particular choice $\lambda=-\alpha/6\mu$ , to ensure
$\delta\rho_{B}< 1$ , the following wave:
\begin{equation}
\delta\rho_{B}(x,t;\mu)={\frac{1}{3}}+{\frac{1}{3}}tanh\Bigg\{-{\frac{\alpha}{6\mu}}
\Bigg[x-\Bigg(c_{s}+{\frac{\alpha}{3}}\Bigg)t\Bigg]\Bigg\}
\label{burgeralsol}
\end{equation}
with the supersonic speed $v_{s}$ :
\begin{equation}
v_{s}=c_{s}+{\frac{\alpha}{3}}
\label{speedsol}
\end{equation}

\begin{figure}[ht!]
\begin{center}
{\label{fig:comp}
\includegraphics[width=0.54\textwidth]{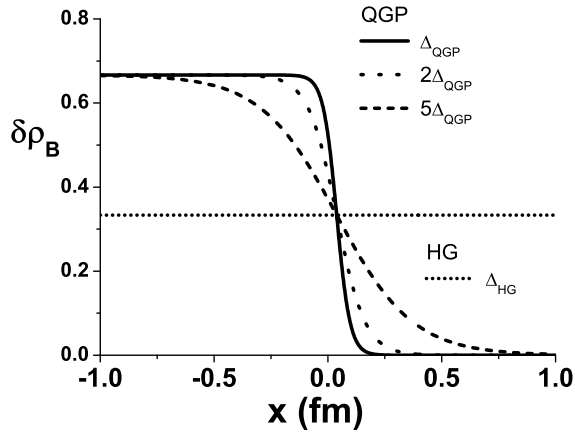}}
\end{center}
\caption{Burgers shock wave solution with different values of the effective viscosity coefficient $\Delta$ for QGP and HG.}
\label{fig1}
\end{figure}

We follow the analysis performed in \cite{pra1,pra2} of the hyperbolic tangent profile (\ref{burgeralsol}).  We fix $t$ for the QGP and HG phases and
obtain the wave profiles shown in  Fig. \ref{fig1} with the parameters of Table \ref{table1}.

\begin{table}[!htbp]
\caption{Parameters of QGP and HG for Fig. \ref{fig1}.}
\vspace{0.3cm}
\centering
\begin{tabular}{cccccc}
\hline
 & $\rho_{0}$ $(fm^{-3})$ & \hspace{0.5cm} ${c_{s}}^{2}$  & \hspace{0.5cm}$\Delta$ $(GeV/fm^{2})$  &
  \hspace{0.5cm} $t$$(fm)$ \\ [0.8ex]
\hline
\hline
$QGP$ & $10\rho_{n}$ & \hspace{0.5cm} $1/3$  &   $0.05$ & $0.05$  \\
\hline
$HG$ &  $1.5\rho_{n}$  & \hspace{0.5cm} $1/3$  & $2$  & $0.05$ \\
\hline
\end{tabular}
\label{table1}
\end{table}

In Fig. \ref{fig1} we observe that in QGP phase the limit of small $\mu$, in (\ref{burgeralsol}) is given by the classical shock wave:
\begin{equation}
\lim_{\mu_{QGP}\to 0} \delta\rho_{B}(x,t;\mu_{QGP})=
\left\{ \begin{array}{ll}
2/3 & \,\,\,\, \textrm{for} \,\,\,\, x/t \,\, < \,\, c_{s}+\alpha /3\\
0 & \,\,\,\, \textrm{for} \,\,\,\, x/t \,\, > \,\, c_{s}+\alpha /3
\end{array} \right.
\label{regbur}
\end{equation}
As in \cite{pra1} we rewrite (\ref{burgeral}) as the following inhomogeneous continuity
equation:
\begin{equation}
{\frac{\partial }{\partial t}}\delta\rho_{B}+
{\frac{\partial }{\partial x}}\Bigg( {c_{s}} \, \delta\rho_{B}+
{\frac{\alpha}{2}} \, {\delta\rho_{B}}^{2}\Bigg)=\mu \, {\frac{\partial^{2} }{\partial x^{2}}}\delta\rho_{B}
\label{burgcontlaw}
\end{equation}
where the limiting process $\mu\to 0$ given by (\ref{regbur}) is a dissipative regularization of the conservation law:
\begin{equation}
{\frac{\partial }{\partial t}}\delta\rho_{B}+
{\frac{\partial }{\partial x}}\Bigg( {c_{s}} \, \delta\rho_{B}+
{\frac{\alpha}{2}} \, {\delta\rho_{B}}^{2}\Bigg)=0
\label{burgcontlawz}
\end{equation}
For the HG phase in Fig. \ref{fig1} the baryon perturbation (\ref{burgeralsol}) is quite different from QGP phase and we have:
\begin{equation}
\delta\rho_{B}(x,t \, ;\mu_{HG}>> \mu_{QGP})={\frac{1}{3}}
\label{nonshockh}
\end{equation}
due to higher background baryon density and higher viscous coefficients.  From these figures we can conclude that, due to the large
viscosity,  shock waves can not be formed in HG. In contrast, they can be easily formed in a QGP. This difference may have phenomenological consequences.

\subsection{Wave packets}

The most important  feature of the nonlinear equation is that there is a
competition between the nonlinear and dissipative coefficients. For some particular parameter choices there will be a balance between the two terms, in which case localized
soliton-like configurations may propagate for long distances. The  numbers employed here suggest that these localized configurations can only survive in the plasma. In order
to illustrate this we will perform the numerical study of the time evolution of a soliton-like pulse. In this case the  initial condition is given by:
\begin{equation}
\delta\rho_{B}(x,t_{0})= A \ sech^{2}\bigg[\frac{x}{B}\bigg]
\label{solitonlike}
\end{equation}
for both (\ref{kdvBxt}) and (\ref{kdvBxtpphase}), with the amplitude $A$ and width $B$ as free parameters.
The viscosity coefficients are extrapolated  from \cite{visc,viscc} and
represented in the effective viscosity coefficient (\ref{efectvis}).
The speed of sound in the HG is calculated by (\ref{relacs}) and is also ${c_{s}}^{2}=1/3$ .

\begin{table}[!htbp]
\caption{Parameters of QGP and HG for Fig. \ref{fig2}.}
\vspace{0.3cm}
\centering
\begin{tabular}{cccccc}
\hline
 & $\rho_{0}$ $(fm^{-3})$ & \hspace{0.5cm} ${c_{s}}^{2}$  & \hspace{0.5cm}$\Delta$ $(GeV/fm^{2})$  &
  \hspace{0.5cm} $B$$(fm)$ \\ [0.8ex]
\hline
\hline
$QGP$ & $10\rho_{n}$ & \hspace{0.5cm} $1/3$  &   $0.05$ & $0.5$  \\
\hline
$HG$ &  $1.5\rho_{n}$  & \hspace{0.5cm} $1/3$  & $2$  & $0.5$ \\
\hline
\end{tabular}
\label{table2}
\end{table}

\begin{figure}[ht!]
\begin{center}
\subfigure[ ]{\label{fig:first}
\includegraphics[width=0.48\textwidth]{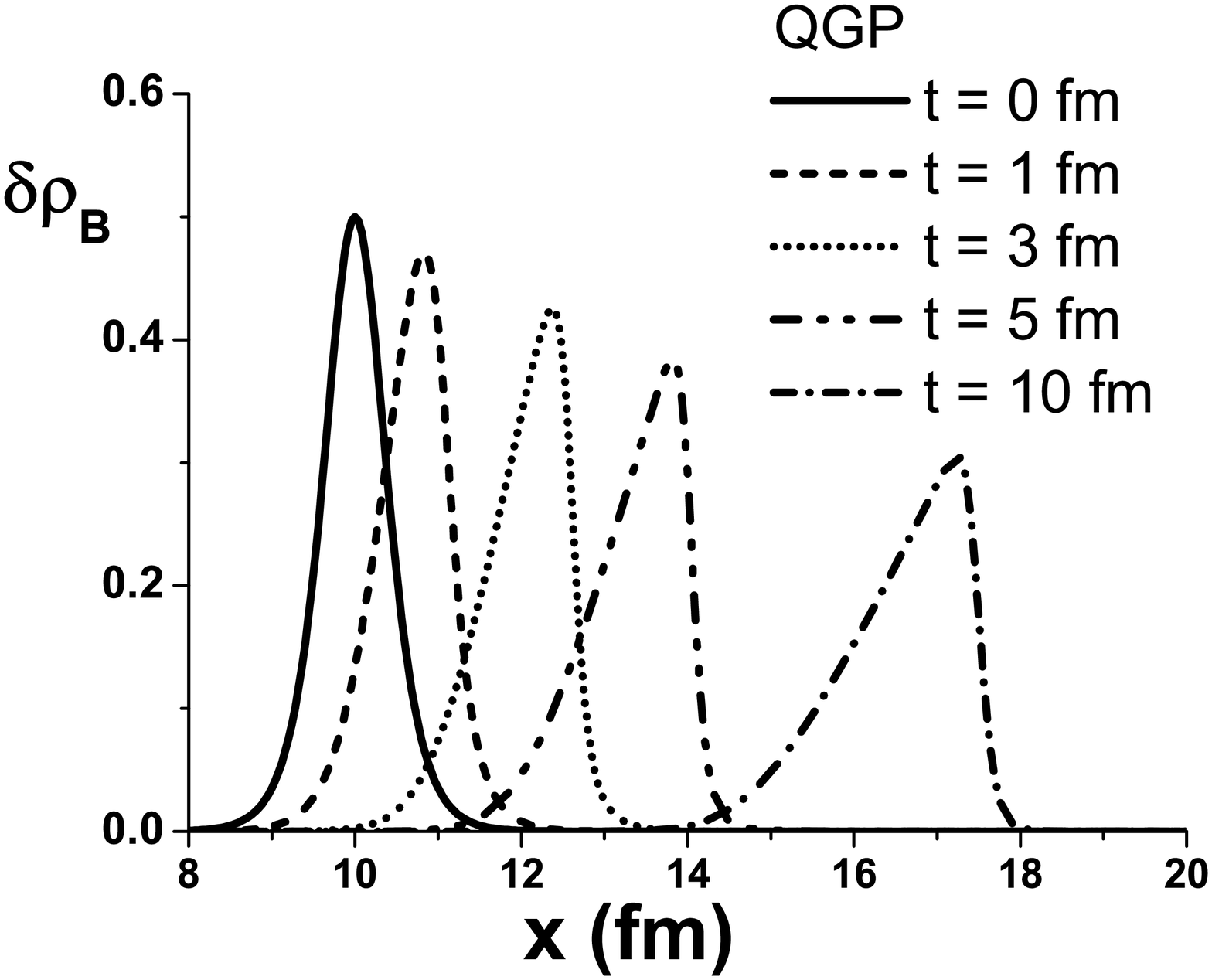}}
\subfigure[ ]{\label{fig:second}
\includegraphics[width=0.48\textwidth]{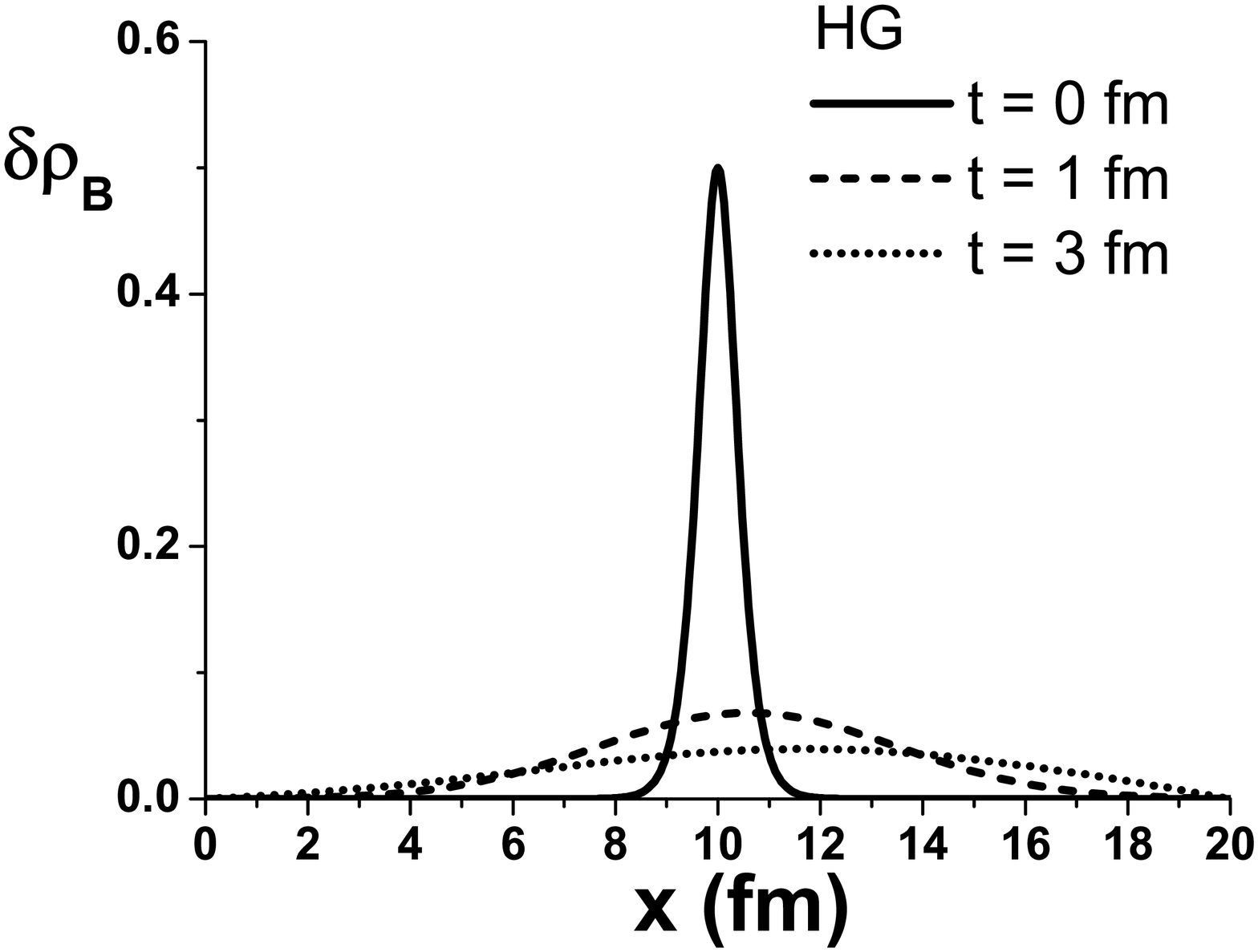}}\\
\subfigure[ ]{\label{fig:third}
\includegraphics[width=0.48\textwidth]{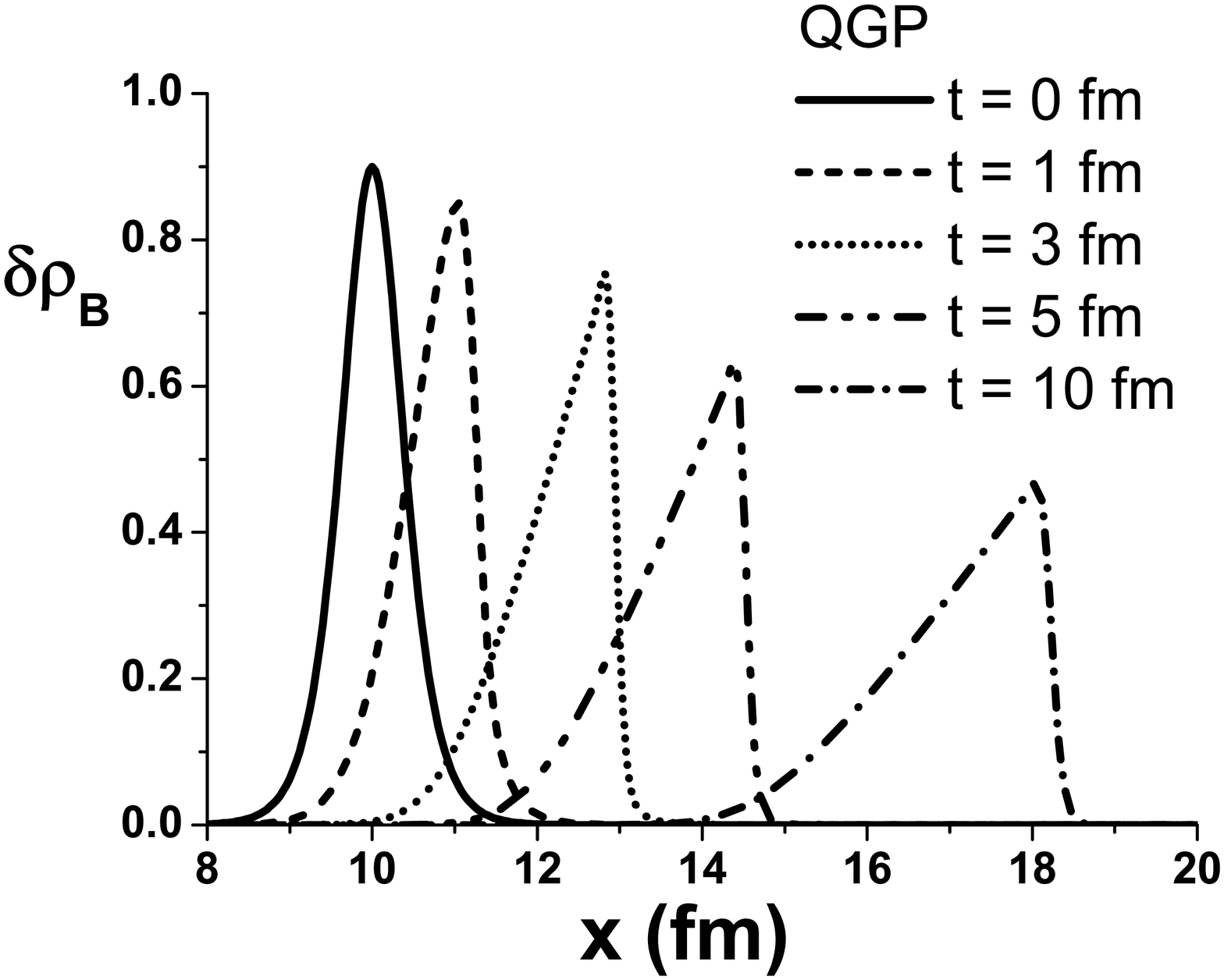}}
\subfigure[ ]{\label{fig:fourth}
\includegraphics[width=0.48\textwidth]{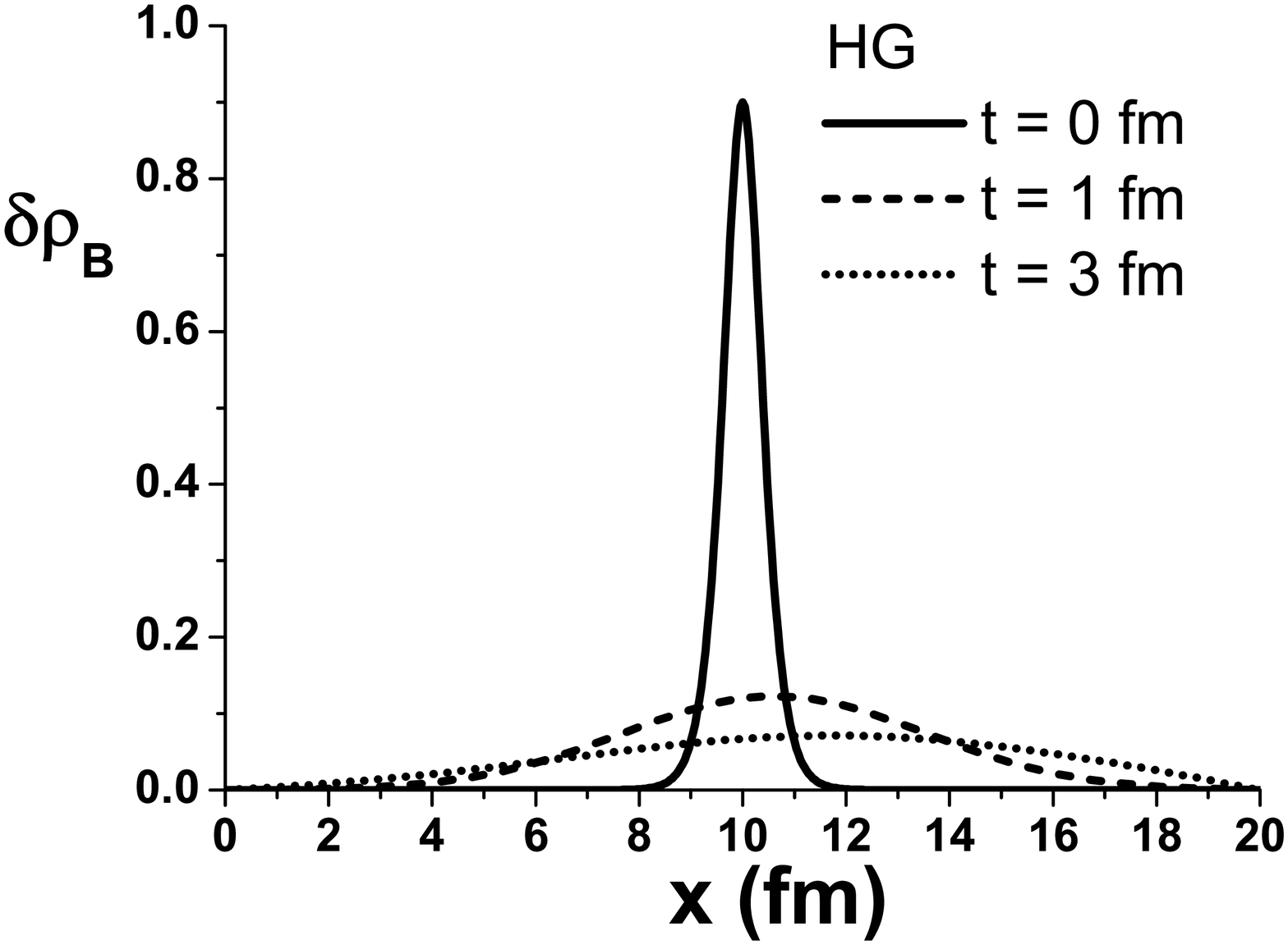}}
\end{center}
\caption{Viscosity effects in hadron and plasma phase.}
\label{fig2}
\end{figure}

In the analysis of the numerical solutions we must carry out a comprehensive  study varying each parameter independently, keeping all others fixed.
In particular, we study how the solution changes when we  change the viscosity.
Such a study was performed in \cite{fgn12} for the numerical solutions of a similar wave equation, derived from the relativistic Navier-Stokes equation
applied to the QGP at high temperatures.  The conclusion was that the most relevant parameter was  the viscosity.
As for the others, we find that they have here the same relative importance as in \cite{fgn12}. In Fig. \ref{fig2}  we compare the hadron gas (on the two right
panels) with the quark gluon plasma (on the two left panels) for the same scales.
To obtain the curves shown in the figure we have used the numbers given in Table II.
Comparing left and right we can observe the strong damping which happens in the
hadron gas, which comes ultimately from the larger values of the viscosity coefficients. Comparing the upper with the lower panels we see that increasing the initial
amplitude of the pulse in both phases does not change much its evolution, damping and slowing down.  In the plasma phase we can also
observe  the formation of a ``wall'' as can be seen in Fig. \ref{fig:first}  and  Fig. \ref{fig:third}.

\section{Conclusions}

In this work we have studied wave propagation in cold and dense matter, both in a hadron gas and in a quark gluon plasma phase.  In deriving wave equations from the
equations of hydrodynamics, we have considered both smaller and larger amplitude waves. The former were treated with the linearization approximation while the latter
were treated with the reductive perturbation method.

Linear waves were obtained by solving an inhomogeneous viscous wave equation and they have the familiar form of
sinusoidal traveling waves multiplied by an exponential damping factor, which depends on the viscosity coefficients. Since these coefficients differ by two orders
of magnitude, even without any numerical calculation we can conclude that, apart from extremely special parameter choices, in contrast to the quark gluon plasma there
will be no linear wave propagation in a hadron gas.

Nonlinear waves were obtained by solving the Burgers equation, which was derived from the equations of hydrodynamics with the RPM. We could find an analytical solution of
the Burgers equation which describes a shock wave. Varying the viscosity parameters in the appropriate range we could conclude that, with the exception of
extremely special parameter choices, there is no shock formation in a hadron gas. If some external agent would try to create a sharp density discontinuity in a
hadronic medium, viscosity would immediately smooth it and wash it out. In a quark gluon plasma, on the other hand, shocks may be formed.
Exactly the same features were observed in the wave packet numerical solutions of the Burgers
equation. In the hadron gas, the wave packet moves very slowly and its amplitude is very quickly reduced. In the quark gluon plasma there is a balance between
nonlinearity and dissipation that prevents the wave from  breaking and from dispersion. Density perturbations may ``survive'' longer in QGP than in HG.

We believe that our work may help in discriminating the quark gluon plasma from the hadron gas. The next and most difficult step is to connect the waves studied here
with observables and plug our analytical studies in realistic numerical simulations of low energy heavy ion collisions or compact stars. Work in this direction is in
progress.

\section{Appendix}

\begin{figure}[ht!]
\begin{center}
{\label{fig:comp}
\includegraphics[width=0.50\textwidth]{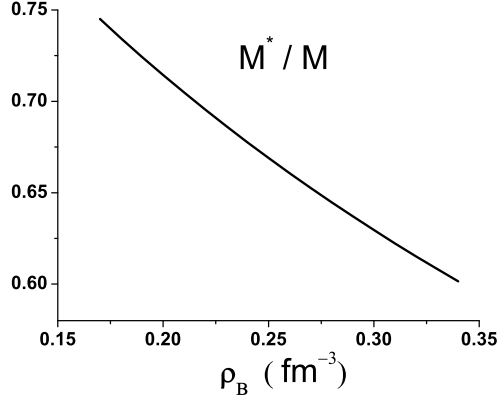}}
\end{center}
\caption{Numerical solution of (\ref{efmass}).}
\label{fig0}
\end{figure}

Inserting (\ref{bard}) into (\ref{preeps}) and performing the $k-$integration we obtain:
$$
\varepsilon={\frac{{g_{V}}^{2}}{2{m_{V}}^{2}}}\,{\rho_{B}}^{2}
+{\frac{{m_{S}}^{2}}{2{g_{S}}^{2}}}(M-M^{*})^{2}
+b{\frac{(M-M^{*})^{3}}{3{g_{S}}^{3}}}
+c{\frac{(M-M^{*})^{4}}{4{g_{S}}^{4}}}
$$
$$
+{\frac{a^{1/3}}{2{\pi}^{2}}}{\rho_{B}}^{1/3}
\sqrt{{\Big(a^{2/3}{\rho_{B}}^{2/3}+{M^{*}}^{2}\Big)}^{3}}
-{\frac{a^{1/3}\,{M^{*}}^{2}}{4{\pi}^{2}}}{\rho_{B}}^{1/3}\sqrt{a^{2/3}{\rho_{B}}^{2/3}+{M^{*}}^{2}}
$$
\begin{equation}
-{\frac{{M^{*}}^{4}}{4\pi^{2}}}\ln\Big(a^{1/3}\,{\rho_{B}}^{1/3}+\sqrt{a^{2/3}{\rho_{B}}^{2/3}+{M^{*}}^{2}}\Big)
+{\frac{{M^{*}}^{4}}{4\pi^{2}}}\ln({M^{*}})
\label{aeps}
\end{equation}
where $a$ is the numerical factor that comes from (\ref{bard}) and is given by $a \equiv (3\pi^{2}/2)$.
Finding the numerical solution of (\ref{efmass}) we obtain the
plot shown in Fig. \ref{fig0} for the ratio $M^{*}/M$, valid only for $\rho_{0} \leq \rho_{B} \leq 2\rho_{0}$, where $\rho_{0}=0.17 \, fm^{-3}$ is the normal nuclear density.
We consider the small parameter \, $0.15 \leq {a^{2/3}{\rho_{B}}^{2/3}}/{{M^{*}}^{2}} \leq 0.36$ \,\, for \,\, $\rho_{0} \leq \rho_{B} \leq 2\rho_{0}$ and perform the following approximations:

$i)$
\begin{equation}
{\frac{a^{1/3}}{2{\pi}^{2}}}{\rho_{B}}^{1/3}\sqrt{{\Big(a^{2/3}{\rho_{B}}^{2/3}+{M^{*}}^{2}\Big)}^{3}}
%$$
%={\frac{a^{1/3}}{2{\pi}^{2}}}{\rho_{B}}^{1/3}\sqrt{\Bigg[{M^{*}}^{2}{\Bigg(1+{\frac{a^{2/3}{\rho_{B}}^{2/3}}{{M^{*}}^{2}}}\Bigg)\Bigg]}^{3}}
%$$
%$$
%={\frac{a^{1/3}}{2{\pi}^{2}}}{\rho_{B}}^{1/3}{M^{*}}^{3}\sqrt{{\Bigg(1+{\frac{a^{2/3}{\rho_{B}}^{2/3}}{{M^{*}}^{2}}}\Bigg)}^{3}}
%\cong{\frac{a^{1/3}}{2{\pi}^{2}}}{\rho_{B}}^{1/3}{M^{*}}^{3}\Bigg[1+{\frac{3a^{2/3}{\rho_{B}}^{2/3}}{2{{M^{*}}^{2}}}} %\Bigg]
%$$
\cong {\frac{a^{1/3}{M^{*}}^{3}}{2{\pi}^{2}}}{\rho_{B}}^{1/3}+
{\frac{3a{M^{*}}}{4\pi^{2}}}{\rho_{B}}
\label{rad1}
\end{equation}

$ii)$
\begin{equation}
-{\frac{a^{1/3}\,{M^{*}}^{2}}{4{\pi}^{2}}}{\rho_{B}}^{1/3}
\sqrt{a^{2/3}{\rho_{B}}^{2/3}+{M^{*}}^{2}}
%$$
%$$
%=-{\frac{a^{1/3}\,{M^{*}}^{2}}{4{\pi}^{2}}}{\rho_{B}}^{1/3}{M^{*}}\sqrt{1+{\frac{a^{2/3}{\rho_{B}}^{2/3}}{{M^{*}}^{2}}}}
%$$
%\begin{equation}
%\cong-{\frac{a^{1/3}\,{M^{*}}^{3}}{4{\pi}^{2}}}{\rho_{B}}^{1/3}\Bigg[1+{\frac{a^{2/3}{\rho_{B}}^{2/3}}{2{M^{*}}^{2}}}  %\Bigg]
\cong -{\frac{a^{1/3}\,{M^{*}}^{3}}{4{\pi}^{2}}}{\rho_{B}}^{1/3}
-{\frac{a{M^{*}}}{8{\pi}^{2}}}{\rho_{B}}
\label{rad2}
\end{equation}

$iii)$
$$
-{\frac{{M^{*}}^{4}}{4\pi^{2}}}\ln\Big(a^{1/3}\,{\rho_{B}}^{1/3}+\sqrt{a^{2/3}{\rho_{B}}^{2/3}+{M^{*}}^{2}}\Big)
$$
\begin{equation}
\cong -{\frac{{M^{*}}^{4}}{4\pi^{2}}}\ln({M^{*}})
-{\frac{a^{1/3}{M^{*}}^{3}}{4\pi^{2}}}{\rho_{B}}^{1/3}
-{\frac{a^{2/3}{M^{*}}^{2}}{8\pi^{2}}}{\rho_{B}}^{2/3}
\label{ln}
\end{equation}
Inserting (\ref{rad1}), (\ref{rad2}) and (\ref{ln}) into (\ref{aeps}) we find:
\begin{equation}
\varepsilon={\frac{{g_{V}}^{2}}{2{m_{V}}^{2}}}\,{\rho_{B}}^{2}
+{\frac{{m_{S}}^{2}}{2{g_{S}}^{2}}}(M-M^{*})^{2}
+b{\frac{(M-M^{*})^{3}}{3{g_{S}}^{3}}}
+c{\frac{(M-M^{*})^{4}}{4{g_{S}}^{4}}}
%$$
%\begin{equation}
+{\frac{5aM^{*}}{8\pi^{2}}}{\rho_{B}}
-{\frac{a^{2/3}{M^{*}}^{2}}{8{\pi}^{2}}}{\rho_{B}}^{2/3}
\label{aepsf}
\end{equation}
Fitting the curve in Fig.\ref{fig0} by a linear parametrization we obtain:
\begin{equation}
{M^{*}}/M\cong 0.9-0.84\, \rho_{B}
\label{linear}
\end{equation}
where $[\rho_{B}]=fm^{-3}$.
Inserting (\ref{linear}) into (\ref{aepsf}) we find:
$$
\varepsilon={\frac{{g_{V}}^{2}}{2{m_{V}}^{2}}}\,{\rho_{B}}^{2}
+{\frac{{m_{S}}^{2}}{2{g_{S}}^{2}}}(0.84 \,M\, \rho_{B}+0.1 \,M)^{2}
+b{\frac{(0.84 \,M\, \rho_{B}+0.1 \,M)^{3}}{3{g_{S}}^{3}}}
+c{\frac{(0.84 \,M\, \rho_{B}+0.1 \,M)^{4}}{4{g_{S}}^{4}}}
$$
\begin{equation}
+{\frac{5a}{8\pi^{2}}}{\rho_{B}}(0.9\,M-0.84\,M\, \rho_{B})
-{\frac{a^{2/3}}{8{\pi}^{2}}}{\rho_{B}}^{2/3}
(0.9\,M-0.84\,M\, \rho_{B})^{2}
\label{aepsmefuseda}
\end{equation}
which can be organized in a power series of $\rho_{B}$ yielding Eq. (\ref{aepsmefusedag}).

\begin{acknowledgments}
The authors are grateful to Jorge Noronha for fruitful discussions. One of the authors (L. G. F. F.) acknowledges Prof. M. McCall for
his hospitality in the Department of Physics and Astronomy at York University.
This work was  partially financed by the Brazilian funding agencies CAPES, CNPq and FAPESP.
\end{acknowledgments}


\begin{thebibliography}{99}

\bibitem{signal}   S. A. Bass, M. Gyulassy, H. Stoecker and W. Greiner,
  J.\ Phys.\ G {\bf 25}, R1 (1999);
  S. Scherer, S. A. Bass, M. Bleicher, M. Belkacem, L. Bravina, J. Brachmann, A. Dumitru and C. Ernst {\it et al.},
  Prog.\ Part.\ Nucl.\ Phys.\  {\bf 42}, 279 (1999);    C. -Y. Wong,
  Nucl.\ Phys.\ A {\bf 681}, 22 (2001).


\bibitem{matsui} T. Matsui and H. Satz,
  Phys.\ Lett.\ B {\bf 178}, 416 (1986).

\bibitem{muller}   P. Koch, B. Muller and J. Rafelski,
Phys.\ Rept.\  {\bf 142}, 167 (1986).

\bibitem{new-qgp}   U. W. Heinz,  arXiv:1304.3634 [nucl-th]; E. Shuryak,
  Prog.\ Part.\ Nucl.\ Phys.\  {\bf 62}, 48 (2009).


\bibitem{qgp-visc}   U. Heinz, C. Shen and H. -C. Song,
  AIP Conf.\ Proc.\  {\bf 1441}, 766 (2012), arXiv:1108.5323.

\bibitem{visc}  A. K. Chaudhuri,   J.\ Phys.\ G: Nucl. Part. Phys. {\bf 39}, 125102 (2012); V. Roy  and A. K. Chaudhuri,  arXiv:1201.4230 [nucl-th].


\bibitem{viscc}  Oleg N. Moroz, arXiv:1301.6670 [hep-th]; arXiv:1112.0277 [hep-th].

\bibitem{visc-eff}    C. Gale, S. Jeon and B. Schenke,
  Int.\  J.\  Mod.\  Phys.\  A  {\bf 28}, 1340011 (2013);
 C. Gale, S. Jeon, B. Schenke, P. Tribedy and R. Venugopalan,
  Phys.\ Rev.\ Lett.\  {\bf 110}, 012302 (2013);
 B. Schenke, S. Jeon and C. Gale,
  Phys.\ Rev.\ C {\bf 85}, 024901 (2012)

\bibitem{kapu} J. I. Kapusta, B. M{\"u}ller, and M. Stephanov, Phys. Rev. C {\bf 85}, 054906 (2012).

\bibitem{shuryak1}   E. Shuryak, Phys. Rev. C {\bf 80}, 054908 (2009).

\bibitem{shuryak2}   P. Staig and E. Shuryak,  Phys.\ Rev.\  C {\bf 84}, 044912 (2011);
                                               Phys.\ Rev.\  C {\bf 84}, 034908 (2011);
                                               J. Phys. G {\bf 38}, 124039 (2011); arXiv:1109.6633.

\bibitem{flor} S. Florchinger and U. A. Wiedemann,  JHEP {\bf 1111}, 100 (2011).

\bibitem{peter}  H. Petersen, G.-Y. Qin, S. A. Bass, and B. Muller, Phys. Rev. C {\bf 82}, 041901 (2010);
                 G.-Y. Qin, H. Petersen, S. A. Bass, and B. Muller, Phys. Rev. C {\bf 82}, 064903 (2010);
                 B. Schenke, S. Jeon, and C. Gale, Phys. Rev. Lett. {\bf 106}, 042301 (2011);
                 Z. Qiu and U. W. Heinz, Phys. Rev. C {\bf 84}, 024911 (2011);
                 Y. Cheng, Y.-L. Yan, D.-M. Zhou, X. Cai, B.-H. Sa, and L. P. Csernai, Phys. Rev. C {\bf 84}, 034911 (2011).
                 R. S. Bhalerao, M. Luzum, and J.-Y. Ollitrault, Phys. Rev. C {\bf 84}, 054901 (2011).

\bibitem{iniflu}  R. P. G. Andrade, F. Grassi, Y. Hama and W. -L. Qian, Phys.\ Lett.\ B {\bf 712}, 226 (2012);
                  R. P. G. Andrade, F. Grassi, Y. Hama and W. -L. Qian, Nucl.\ Phys.\ A {\bf 854}, 81 (2011).

\bibitem{fodor} Y. Aoki, G. Endr\H{o}di, Z. Fodor, S. D. Katz and K. K. Szab\'o, Nature {\bf 443} (2006) 675.

\bibitem{fair}  Johann M. Heuser for the CBM collaboration, Nucl. Phys. A {\bf 830}, 563c (2009);
                B. Friman {\it et al.}, ``The CBM Physics Book'', Lect. Notes Phys. 814, Springer-Verlag Berlin
                Heidelberg (2010).

\bibitem{raha}  G. N. Fowler, S. Raha, N. Stelte and R. M. Weiner,  Phys. Lett. B {\bf 115}, 286 (1982).

\bibitem{fn06} D. A. Foga\c{c}a and  F. S. Navarra, Phys. Lett. B {\bf 639}, 629 (2006).

\bibitem{fn07}  D. A. Foga\c{c}a and  F. S. Navarra, Phys. Lett. B {\bf 645}, 408 (2007);
                Nucl. Phys. A  {\bf 790}, 619c (2007).

\bibitem{fn07a}  D. A. Foga\c{c}a and F. S. Navarra,
  Int.\ J.\ Mod.\ Phys.\ E {\bf 16}, 3019 (2007).


\bibitem{ffn09} D. A. Foga\c{c}a, L. G. Ferreira Filho and  F. S. Navarra,
                Nucl. Phys. A {\bf 819}, 150 (2009).


\bibitem{ffn10} D. A. Foga\c{c}a, L. G. Ferreira Filho and  F. S. Navarra,
                Phys. Rev. C {\bf 81}, 055211 (2010).


\bibitem{ffn11} D. A. Foga\c{c}a, F. S. Navarra and L. G. Ferreira Filho,
                Phys.\ Rev.\  D {\bf 84}, 054011 (2011).


\bibitem{fgn12} D. A. Foga\c{c}a,  F. S. Navarra and  L. G. Ferreira Filho,
                Nucl. \ Phys.  A {\bf 887}, 22  (2012).

\bibitem{rand1} J. Randrup, Phys. Rev. C {\bf 82}, 034902 (2010).

\bibitem{rand2} J. Steinheimer and J. Randrup, Phys. Rev. C {\bf87}, 054903 (2013).

\bibitem{shokov1} V. V. Skokov and D. N. Voskresensky, Nucl. Phys. A {\bf 828}, 401 (2009).

\bibitem{kapusta} L. P. Csernai and J. I. Kapusta, Phys. Rev. D {\bf 46}, 1379 (1992).

\bibitem{roma} P. Romatschke, Int. J. Mod. Phys. E {\bf 19}, 1 (2010).
					
\bibitem{wein} S. Weinberg,``Gravitation and Cosmology'', New York: Wiley, (1972).

\bibitem{land} L. Landau and  E. Lifchitz, ``Fluid Mechanics'',
                Pergamon Press, Oxford, (1987).				

\bibitem{hidro} J. Y. Ollitrault, {\sl Eur. J. Phys.} {\bf 29}, 275 (2008);
                  arXiv:0708.2433 [nucl-th]

\bibitem{davidson} R. C. Davidson, ``Methods in Nonlinear Plasma Theory'',
                   Academic Press,   New York and London, (1972).
			
\bibitem{rpm} H. Washimi and T. Taniuti,  Phys. Rev. Lett. {\bf 17}, 996 (1966).

\bibitem{leblond} H. Leblond, J. Phys. B: At. Mol. Opt. Phys. {\bf 41}, 043001 (2008).

\bibitem{loke} Lokenath Debnath, ``Nonlinear Partial Differential Equations for Scientists and Engineers'', third edition,
               Birkh\"auser, USA, (2011).

\bibitem{nrev}  D. A. Foga\c{c}a,  F. S. Navarra and  L. G. Ferreira Filho, `` Solitons:
Interactions, Theoretical and Experimental Challenges and
Perspectives'' (Nova Science Publishers, New York, 2013), arXiv:1212.6932.

\bibitem{fng}  D. A. Foga\c{c}a, F. S. Navarra and L. G. Ferreira Filho, Comm. Nonlin. Sci. Num. Sim. {\bf 18}, 221 (2013).

\bibitem{lac} A. Lacour, J. A. Oller and U.-G. Meissner, Annals of Physics {\bf 326}, 241 (2011).

\bibitem{furn} R. J. Furnstahl,  Lect. Notes Phys. {\bf 641}, 1 (2004);
               B. D. Serot,   Int. J. Mod. Phys. {\bf A19S1}, 107 (2004)
                     and  references therein.

\bibitem{serot} Brian D. Serot and John Dirk Walecka,
                 Adv. in Nucl. Phys. {\bf 16}, 1 (1986).	

\bibitem{fuku} K. Fukushima and C. Sasaki, arXiv:1301.6377.


\bibitem{dp} A. L. Espindola and D. P. Menezes,
			  Phys. Rev.  {\bf C65}, 045803 (2002);
              A. M. S. Santos and D. P. Menezes,
			   Braz. J.  Phys. {\bf 34}, 833 (2004).

\bibitem{osalemao} H. Peter and R. Schlichenmaier,
http://www3.kis.uni-freiburg.de/~schliche/lectures\_files/hydro.pdf

\bibitem{stvis1} N. Antar, International Journal of Engineering Science {\bf 40}, 1179 (2002).

\bibitem{stvis2} R. Saeed and A. Shah, Phys. Plasmas {\bf 17}, 032308 (2010).

\bibitem{pra1} M. A. Hoefer, M. J. Ablowitz, I. Coddington, E. A. Cornell, P. Engels and V. Schweikhard, Phys. Rev. A {\bf 74}, 023623  (2006).

\bibitem{egypt} T. S. El-Danaf and M. A. Ramadan, Open Appl. Math. J. {\bf I}, 1 (2007).

\bibitem{pra2} M. Kulkarni and A. G. Abanov, Phys. Rev. A {\bf 86}, 033614  (2012).


\end{thebibliography}
\end{document}